\documentclass[a4paper,11pt]{article}
\usepackage{jheppub}
\usepackage{amsmath,amssymb,latexsym,bbm}
\usepackage{braket}
\usepackage{bm}
\usepackage{booktabs}
\usepackage{setspace}
\usepackage{hepunits}
\usepackage[svgnames]{xcolor}
\usepackage{graphicx}

\usepackage{mciteplus}

\usepackage{tikz}
\usepackage{graphicx}
\usetikzlibrary{decorations.pathreplacing}
\usetikzlibrary{decorations.pathreplacing,calligraphy}
\usepackage{graphicx}

\newcommand{\nn}{\nonumber}
\newcommand{\nd}{\mathrm{d}}
\newcommand{\nD}{\mathrm{D}}
\newcommand{\nT}{\mathrm{T}}
\newcommand{\nW}{\mathrm{W}}
\newcommand{\nds}{\hat{\mathrm{d}}}

\newcommand{\mG}{\mathcal{G}}
\newcommand{\spdot}{\!\cdot\!}
\newcommand{\nG}{\mathrm{G}}
\newcommand{\ep}{\epsilon}

\newcommand{\bea}{\begin{eqnarray}}
	\newcommand{\eea}{\end{eqnarray}}
\newcommand{\be}{\begin{equation}}
	\newcommand{\ee}{\end{equation}}
\newcommand{\nnn}{\nonumber\\}
\allowdisplaybreaks

\def\eref#1{(\ref{#1})}

\DeclareMathOperator{\Res}{Res}

\usepackage{color}  
\usepackage{xcolor}
\definecolor{ceil}{rgb}{0.57, 0.63,0.81}

\title{Notes on Selection Rules of Canonical Differential Equations and Relative Cohomology}
 
\author[a]{Jiaqi Chen}
\author[a,b]{Bo Feng}

\affiliation[a]{Beijing Computational Science Research Center, Beijing 100084, China}
\affiliation[b]{Peng Huanwu Center for Fundamental Theory, Hefei, Anhui 230026, China}

\emailAdd{jiaqichen@csrc.ac.cn}
\emailAdd{fengbo@csrc.ac.cn}

\abstract{
 We give an explanation of the $\mathrm{d}\log$-form of the coefficient matrix of canonical differential equations
using the projection of ($n$+1)-$\mathrm{d}\log$ forms onto $n$-$\mathrm{d}\log$ forms. This projection
is done using the leading-order formula for intersection numbers. This formula gives a simple way to compute the coefficient matrix.
When combined with the relative twisted cohomology, redundancy in computation using the regulator method can be avoided. 

}

\begin{document}
\maketitle

\section{Introduction}

For the study of theories and phenomenology in high energy physics,  perturbative quantum field theory plays a crucial role. One of its central tasks is to compute Feynman integrals. For Feynman integrals,  linear relationships can be established through Integration-By-Parts (IBP) \cite{Chetyrkin:1981qh}, thereby expressing integrals within the same function family as  linear combinations of a chosen finite set of integrals. These selected integrals are referred to as master integrals, and this process is known as the IBP reduction. Subsequently, the computation in perturbative field theory is transformed into completing the reduction and calculating the master integrals. 
Based on IBP, one can take partial derivatives of master integrals and then use IBP reduction to express the resulting integrals  in terms of master integrals. This process leads to  first-order differential equations satisfied by the master integrals \cite{Kotikov:1990kg,Kotikov:1991pm,Gehrmann:1999as,Bern:1993kr}. To simplify the
computations, {\bf canonical differential equations (CDE)} method \cite{Henn:2013pwa} are developed. It says  that for many cases, when    the master integrals  are appropriately chosen, the differential equations will be transformed into $\nd \log$-form proportional to 
$\ep$, i.e., 
\begin{align}
&\nd f_I = \big( \nd \Omega \big)_{IK} f_K \, , \quad
\big( \nd \Omega \big)_{IK} = \ep \sum_i \text{C}^{(i)}_{IK} ~ \nd \log \text{W}^{(i)}(\bm{s})  \, .
\end{align}
Then each order of $\ep$ of the master integrals can be iteratively resolved as iterative integration of a series of $\nd \log \text{W}^{(i)}(\bm{s})$ functions, which leads to  multi-polylogarithm \cite{Chen:1977oja, Goncharov:1998kja}. The $\text{W}^{(i)}(\bm{s})$'s are called symbol letters and they contain the information on the analytic structure of Feynman integrals. The complete set of letters is defined as symbol  alphabet. Symbol has been studied in various researches \cite{Caron-Huot:2011dec, Golden:2013xva, Panzer:2014caa, Dennen:2015bet, Caron-Huot:2016owq, Mago:2020kmp, Abreu:2021vhb, Gong:2022erh, Yang:2022gko, He:2021non, He:2021eec, He:2022tph, He:2023qld, Arkani-Hamed:2017ahv, Abreu:2017enx, Abreu:2017mtm, Chen:2022fyw, Dlapa:2023cvx,Jiang:2024eaj} and could be used for bootstrap \cite{Gaiotto:2011dt, Dixon:2011pw, Dixon:2011nj, Brandhuber:2012vm, Dixon:2013eka, Dixon:2014voa, Dixon:2014iba, Drummond:2014ffa, Dixon:2015iva, Caron-Huot:2016owq, Dixon:2016apl, Dixon:2016nkn, Li:2016ctv, Almelid:2017qju, Chicherin:2017dob, Henn:2018cdp, Drummond:2018caf, Caron-Huot:2019vjl, Caron-Huot:2020bkp, Dixon:2020cnr, Dixon:2020bbt, Guo:2021bym, He:2021eec, Dixon:2022rse, Dixon:2022xqh}.  Information of symbol is encoded in the coefficient matrix $\big( \nd \Omega \big)_{IK}$ of CDE. 

In the past decade, CDE has been the most crucial technique for analytically computing Feynman integrals. People have developed many methods to "appropriately choose the master integral" to get CDE,  such as $\nd \log$-form and the closely related leading singularity analysis \cite{Henn:2013pwa, Chen:2020uyk, Chen:2022lzr, Chicherin:2018old, Bern:2014kca, Henn:2020lye, Dlapa:2021qsl}, which are also inspired by previous work such as \cite{Arkani-Hamed:2010pyv, Arkani-Hamed:2012zlh}. There are also some automatic packages with other algorithms such as \cite{Dlapa:2020cwj, Lee:2020zfb, Prausa:2017ltv, Gituliar:2017vzm, Meyer:2017joq, Meyer:2016slj}. People found that if one can construct $\nd \log$-from integrands as master integrals,  the differential equations of this system will become CDE automatically in practice. Baikov representation \cite{Baikov:1996iu} shows advantage in such construction \cite{Chicherin:2018old,Chen:2020uyk, Chen:2022lzr}. However, the origin of CDE is not clear. In the last several years, a mathematical tool called "intersection theory" was introduced to Feynman integral and developed in \cite{Mastrolia:2018uzb, Frellesvig:2019uqt, Mizera:2019ose, Matsubara-Heo:2020lzo, Chestnov:2022xsy, Weinzierl:2020xyy, Chen:2020uyk,  Chen:2022lzr, Jiang:2023oyq, Jiang:2023qnl, Crisanti:2024onv, Frellesvig:2020qot, Mizera:2019vvs, Chestnov:2022alh, Caron-Huot:2021xqj, Caron-Huot:2021iev, Giroux:2022wav, De:2023xue,Duhr:2024rxe,Chen:2023kgw,Brunello:2024tqf}. It could be used as a reduction method equivalent to IBP method.  Recently, people successfully apply it, together with companion tensor algebra, to reduce the 2-loop 5-point Feynman integral family  \cite{Brunello:2024tqf}. In one of the previous work \cite{Chen:2023kgw}, how CDE emerges from $\nd \log$-from integrand has been partially understood by using intersection theory, especially the method of computing intersection numbers from higher-order partial differential equations \cite{Chestnov:2022xsy}. Furthermore, selection rules of the coefficient matrix of CDE could be given, including which element of the matrix is non-zero and which letter could appear.
For these computations, all people need are two universal formulas, i.e., formulas for the leading-order (LO) contribution and next-to-leading-order (NLO) contribution of the intersection number.

In this paper, we are going to improve the selection rules of the coefficient matrix  of CDE, mainly in two aspects. Firstly we observe that for $\nd \log$-form basis,  one can easily transform the differential action on $n$-$\nd \log$-form basis to a $(n+1)$-$\nd \log$-form. 
This rewriting is very useful since the usage of the formula of  NLO contribution of intersection number could be avoided when computing the coefficient matrix of CDE. Furthermore, it is easy to see that the coefficient matrix  is nothing but the $\nd \log$-form coefficient when projecting $(n+1)$-$\nd \log$-form to $n$-$\nd \log$-form.  Secondly, in the previous work \cite{Chen:2023kgw}, factors with integer powers such as propagators are handled by adding regulators that ultimately need to be taken to zero. This makes both the computational process as well as the selection rules more complicated. In contrast, relative twisted cohomology of the dual form \cite{Caron-Huot:2021xqj, Caron-Huot:2021iev, Giroux:2022wav, De:2023xue} in intersection theory could be a natural language to deal with these cases, allowing to avoid regulator from the beginning, and thus obtaining the simpler selection rules.

The organization of this paper is as follows. In section \ref{sec:dlog}, we recall the intersection theory with regulator and the previous version of the CDE selection rule, but from a new perspective, i.e., the $\nd \log$ projecting and with only LO formula, instead of LO and NLO in \cite{Chen:2023kgw}. An important technical point is given in the subsection \ref{sec:2.4}, where we have carefully discussed the factorization of poles, including the understanding of relations between them and integration contours and regions. Then, with the preparation of section \ref{sec:dlog} we can smoothly go  to the computation of intersection number with dual form in relative cohomology in section \ref{sec:RC}. We will introduce this mathematical tool in an easier practice way. Then the improved CDE selection rules are presented. In section \ref{sec:eg1} and \ref{sec:eg2}, we show two examples and compare the computing processes of two methods, i.e., with  regulator and using relative cohomology. From the comparison one can see the simplification of latter. Finally, a summary is given in section \ref{final}.


\section{$\nds \log$-form differential equations from projecting $\mathrm{D} \log$ to $\nd \log$ } \label{sec:dlog}

\subsection{Intersection theory} \label{sec:intnum}
The Feynman integrals in the Baikov representation are  functions of the form 
\begin{equation}
	I[u,\varphi] \equiv \int  u \, \varphi \,, 
\end{equation}
where
\begin{align}
&\varphi \equiv \hat{\varphi}(\bm{z}) \bigwedge_j \nd z_j =\frac{Q(\bm{z})}{\left(\prod_k D_k^{a_k} \right) \left(\prod_i P_i^{b_i} \right)} \bigwedge_j \nd z_j \, , \nn\\
    &u=\prod_i \left[P_i(\bm{z})\right]^{\beta_i} \,,
    \quad \quad  a_k, b_j\in \mathbb{N}  \, ,
    \label{eq:baikov}
\end{align}
The propagators are denoted by $\bm{z}=(z_1,\ldots,z_n)$. 
The polynomials $D_k(\bm{z})$ are denominators with integer power $a_k$, usually are propagators, while $P_i(\bm{z})$ are denominators with complex powers $\beta_i-b_i$ in total, typically are Gram determinants $\nG(\bm{q})\equiv\det(q_i \cdot q_j)$ of loop and external momenta. The numerator $Q(\bm{z})$ is an arbitrary polynomial of $\bm{z}$. In this section, we are supposed to use a regulator to deal with integer power denominators. Then 
\begin{align}
u=\prod_i \left[P_i(\bm{z})\right]^{\beta_i} \prod_k D_k^{\delta_k}
\end{align}
where  regulators $\delta_k$ are kept in the computation, which will be  taken to be zero in the final result. In the next section, we will discuss how to avoid regulators from the beginning of the computation.

Despite traditional IBP reduction, one can also reduce integrals in such an integral family to master integrals via intersection theory.
To do so, one needs to define IBP-equivalence classes of cocycles \cite{Mastrolia:2018uzb, Frellesvig:2019uqt, Mizera:2019ose, Chestnov:2022xsy} first:
\begin{equation}
\bra{\varphi_L} \equiv \varphi_L \sim \varphi_L + \sum_i \nabla_i \xi_i \,, \, \quad \nabla_{i} = dz_i \wedge (\partial_{z_i} + \hat{\omega}_i) \, , \quad \hat{\omega}_i \equiv \partial_{z_i} \log(u) . \label{eq:braket}
\end{equation}
The dual form in dual space is important in this paper and we will discuss  it more carefully later. At this moment, let us roughly regard the dual space as being consisting of equivalence classes $\ket{\varphi}$ of integrals $I[u^{-1},\varphi]$.
Then, the intersection number is given by
\begin{equation}
	\braket{\varphi_L|\varphi_R} = \sum_{\bm{p}} \Res_{\bm{z} = \bm{p}} \left(\psi_L \varphi_R\right) \, , \quad \nabla_1 \cdots \nabla_n \psi_L = \varphi_L \,, \label{eq:intnum}
\end{equation}
where  $\bm{p}$ are isolated intersection points of  $n$  hypersurfaces belonging to $\mathcal{B}=\{P_1=0,\infty,\cdots, D_1=0,\infty,\cdots\}$ ($P_1=0,\infty$ means $P_1=0, P_1=\infty$ ).

With intersection number as the IBP-invariant inner product, the integral reduction becomes entirely a projection in vector space. For example, to reduce $f_0=\int u \varphi_0$ to master integrals $f_I=\int u \varphi_I$
\begin{align}
f_0=\sum_{I=1}^n c_I f_I
\end{align}
$c_I$ can be calculated via
\begin{align}
c_I=\sum_J \braket{\varphi_0|\varphi_J'}\left(\eta^{-1}\right)_{JI} \,, \quad \eta_{IJ}\equiv \braket{\varphi_I|\varphi_J'} \, .
\end{align}
(Dual basis $\varphi_I'$ are not necessarily equal to $\varphi_I$.)


\subsection{$\nd \log$ projection of CDE} \label{sec:intnum}

For CDE considered in this paper,  $\beta_i$ in \eref{eq:baikov} 
 are proportional to $\epsilon$, and  integrands $\varphi_I$ of master integrals $f_I$ are supposed to the $n$-$\nd \log$-form
\begin{align}
\varphi_I = \bigwedge_{j} \nd \log \nW^{(I)}_j(\bm{z}),
\end{align}
which typically has  two types of building blocks:
\begin{align}
	&\nd \log(z-c) = \frac{d z}{z-c}\,, \nn \\
	&\nd \log(\tau[z,c;c_{\pm}]) = \frac{\sqrt{(c-c_+)(c-c_-)}d z}{(z-c)\sqrt{(z-c_+)(z-c_-)}} \,,\nn\\
	&  \tau[z,c;c_{\pm}]  \equiv \frac{\sqrt{c-c_+} \sqrt{z-c_-}+\sqrt{c-c_-} \sqrt{z-c_+}}{\sqrt{c-c_+}
		\sqrt{z-c_-}-\sqrt{c-c_-} \sqrt{z-c_+}}\,,\label{eq:dlog_basis}
\end{align}
where we denote the  first type as ``rational-type'' and the second as ``sqrt-type''.
For convenience and distinction, we denote the differentiation over integration variables  $\bm{z}$ as $\nd$, the differentiation over  arbitrarily selected one parameter such as a kinetic parameter or mass  as  $\nds$, and
\begin{equation}
\nD = \nd+\nds
\end{equation}
Differential equations are given by the IBP reduction of $\partial_s \bm{f}$ (s could be  any selected parameter whose corresponding total derivative is denoted as $\nds$ as we mentioned)
\begin{equation}
\nds \bm{f} = \Omega_s . \bm{f} ~ \nds s.
\end{equation}
Carrying out the computation we have 
\begin{align}
\nds f_I &= \int \nds \left( u \bigwedge_j \nd \log \nW^{(I)}_j(\bm{z}) \right) \nn\\
&= \int u ~\nds \log u \bigwedge_j \nd \log \nW^{(I)}_j
+ \int u \sum_k \left[  (-1)^k \left(\nd \wedge \nds \log \nW^{(I)}_k \right)\bigwedge_{j\neq k} \nd \log \nW^{(I)}_j \right] \nn\\
&=\int  u ~\nds \log u \bigwedge_j \nd \log \nW^{(I)}_j
+ \int u ~\nd \log u \wedge \sum_k  \left[ (-1)^{k+1} \left( \nds \log \nW^{(I)}_k \right)\bigwedge_{j\neq k} \nd \log \nW^{(I)}_j \right] \nn\\
&=\int u~\nD \log u \bigwedge_j \nD \log \nW^{(I)}_j  \label{eq:Dlogvarphi}
\end{align}
Denote $\nds f_I\equiv\int u \dot{\varphi}_I$, we have 
\begin{align}
\dot{\varphi}_I= \nD \log u \bigwedge_j \nD \log \nW^{(I)}_j~~~\label{tot-D}
\end{align}
Obviously, they are n-$\nd\log$-form with $\nds \log$ coefficient. Result \eref{tot-D}
is important as we will show that in the calculation of  matrix $\nds \Omega$ (here dual basis are selected to be the same as original one)
\begin{align}
    &\bra{\dot{\varphi}_I}  = \big( \nds \Omega \big)_{IJ} \bra{\varphi_J} \,,  \nn\\
    &\big( \nds \Omega \big)_{IK} = \braket{\dot{\varphi}_I|\varphi_J} \big( \eta^{-1} \big)_{JK} \, ,
    \label{eq:intnumDE}
\end{align}
people only need to compute the so called leading order (LO) contribution to intersection number. This will lead to a great simplification. In the next two subsections,  details of related mathematical techniques of such computations will be presented.



\subsection{LO contribution of intersection number}

The computation of intersection numbers involves multivariate residues. For fractional polynomials,  multivariate residues can be computed by transformation law, and global residue can be calculated via the Bezoutian method \cite{Zhang:2016kfo, Larsen:2017aqb}. Since $\psi_L$ in the computation of intersection number usually is not polynomial,  for keeping more information explicitly,  we choose to compute multivariate residue \eref{eq:intnum} directly by solving $\psi$ in multivariate Laurent expansion \cite{Chestnov:2022xsy}.

 However, multivariate residue and Laurent expansion are highly non-trivial and usually cannot be computed variable-by-variable. For example, the Laurent expansion of $1/(z_1 (z_1+z_2))$ depends on the order of the expansion variables. 
To overcome this, one can factorize the poles first, then the Laurent expansion is legal and the high order differential equation of $\psi_L$ can be solved using the expansion. This method is developed in  \cite{Chestnov:2022xsy, Chen:2023kgw} and we are going to  give more discussions about  it here. Factorization also transforms the n-variable residue problem into n one-variable residue problem, whose computations become  trivial.

A further technical difficulty is when the pole $\bm{p}$ is degenerate and thus also non-factorized. To deal with it, 
one needs to involve several regions. Each region has different factorization transformations and different residue contributions. 
As indicated in \cite{Chestnov:2022xsy}, this transformation  likes the one  applied in sector decomposition \cite{Binoth:2000ps, Binoth:2003ak, Binoth:2004jv, Heinrich:2008si}.

 We denote all factorization transformations as $\nT^{(\alpha)}: z_i \to f^{(\alpha)}_i(\bm{x}^{(\alpha)})$, and the corresponding pole after transformation as $\bm{\rho}^{(\alpha)}=\{\rho^{(\alpha)}_1,\rho^{(\alpha)}_1,\cdots,\rho^{(\alpha)}_n\}$. Around a factorized pole, an $n$-form $\varphi$ can be Laurent-expanded safely  as
\begin{equation}
  \varphi= \sum_{\bm{b}} \varphi^{(\bm{b})}\,, \quad  \varphi^{(\bm{b})} = C^{(\bm{b})} \bigwedge_i  \left[ x_i^{(\alpha)} - \rho_i^{(\alpha)} \right]^{b_{i}} \nd x^{(\alpha)}_i \,,  \label{eq:phib}
\end{equation}
where the powers $\bm{b}=(b_1,\ldots,b_n)$. The u could be written as 
\begin{equation}
  \nT^{(\alpha)}\left[u \right]\equiv   u(\nT^{(\alpha)}[\bm{z}]) = \bar{u}_\alpha(\bm{x}^{(\alpha)}) \prod_i \left[ x_i^{(\alpha)} - \rho_i^{(\alpha)} \right]^{\gamma_i^{(\alpha)}} ,
    \label{eq:degenerate-1}
\end{equation}
These remaining hypersurfaces in $\bar{u}_\alpha(\bm{x}^{(\alpha)})$  will not intersect at the point $ \bm{\rho}^{(\alpha)}$, so $\bar{u}_\alpha(\bm{\rho}^{(\alpha)})\neq 0$. 
Thus the leading term of u around the pole is
\begin{equation}
    u(\nT^{(\alpha)}[\bm{z}])\big|_{\bm{x}^{(\alpha)} \to \bm{\rho}^{(\alpha)}} = \bar{u}_\alpha(\bm{\rho}^{(\alpha)}) \prod_i \left[ x_i^{(\alpha)} - \rho_i^{(\alpha)} \right]^{\gamma_i^{(\alpha)}} ,
    \label{eq:degenerate}
\end{equation}
We  define  $\gamma^{(\alpha)}_i$ as the {\bf hypersurface-power} for each variable, where $\alpha$  corresponds to the  transformation  $\nT^{(\alpha)}$.

After the above transformation, the intersection number becomes
\begin{align}
&\braket{\varphi_L|\varphi_R}     = \sum_{\alpha} \Res_{\bm{\rho}^{(\alpha)}} \nT^{(\alpha)}\left[ \psi_L \varphi_R\right] \nn\\
&\quad = \sum_{\alpha} \Res_{\bm{\rho}^{(\alpha)}} \left[ \left( \sum_{\bm{b}_L}  \nabla_1^{-1} \cdots \nabla_n^{-1} \varphi^{(\bm{b}_L)}_L  \right) \sum_{\bm{b}_R}\varphi^{(\bm{b}_R)}_R  \right] \, . \label{eq:intnumexpand}
\end{align}
As has been discussed in \cite{Chen:2023kgw}, only when there exist non-zero terms $\varphi^{(\bm{b}_L)}_L$ and $\varphi^{(\bm{b}_R)}_R$ in their expansion which satisfy $b_{L,i}+b_{R,i} \leq -2$ for all $i$, intersection number gets non-zero contribution from such pairs. When $b_{L,i}+b_{R,i} = -2$, we say it gives a {\bf LO contribution} to the intersection number. 
A special case  is the intersection number of $\nd \log$-form. Since the $\nd \log$-form has only multivariate simple poles, all terms in its expansion (after factorization) have all $b_i\geq -1$. Hence, all non-zero contributions come from terms with $\bm{b}_L=\bm{b}_R=\bm{-1}=\{-1,-1,\cdots,-1\}$. Thus the formula of LO contribution can be easily read  as 
\begin{align}
\Res_{\bm{\rho}^{(\alpha)}} \nT^{(\alpha)}\left[  \left( \nabla_1^{-1} \cdots \nabla_n^{-1}  \varphi_L \right)  \varphi_R \right] &= \frac{\Res_{\bm{\rho}^{(\alpha)}} \nT^{(\alpha)}\left[\varphi_L \right]  \times  \Res_{\bm{\rho}^{(\alpha)}} \nT^{(\alpha)}\left[\varphi_R \right]  }{ \prod_{i} \Res_{\rho^{(\alpha)}_i} \partial_{x_i^{(\alpha)}} \log\left(\nT^{(\alpha)}\left[u \right] \right) \nd x_i^{(\alpha)} } \nn\\
& =  \frac{C_L^{(\bm{b}_L)} C_R^{(\bm{b}_R)}}{\bm{\gamma}^{(\alpha)}} \, , \quad \quad \bm{\gamma}^{(\alpha)}= \prod_i \gamma_i^{(\alpha)} \label{eq:intnumdlog}
\end{align}
Let us give some explanations of the result \eref{eq:intnumdlog}. First the term 
$
\partial_{x_i^{(\alpha)}} \log\left(\nT^{(\alpha)}\left[u \right] \right) \nd x_i^{(\alpha)} $
comes from 
\begin{align}
\nT^{(\alpha)}[ \nabla_i] = \nd x_i^{(\alpha)} \wedge \left(\partial_{x_i^{(\alpha)}} +  \partial_{x_i^{(\alpha)}} \log\left(\nT^{(\alpha)}\left[u \right]\right) \nd x_i^{(\alpha)} \right)
\end{align}
which keeps the structure of the commutator
\begin{align}
\left[\nT^{(\alpha)}[\nabla_i],\nT^{(\alpha)}[\nabla_j]\right] = \left[\nabla_i,\nabla_j\right] = 0 \,.
\end{align}
Secondly after solving high-order differential equation 
\begin{align}
	\prod_i \nT^{(\alpha)}[\nabla_i] \nT^{(\alpha)}\left[\psi_L \right] = \nT^{(\alpha)}\left[\varphi_L \right] 
\end{align}
in Laurent series expansion, we pick out the coefficient of the order that will contribute and get 
the part $
\frac{\Res_{\bm{\rho}^{(\alpha)}} \nT^{(\alpha)}\left[\varphi_L \right]    }{ \prod_{i} \Res_{\rho^{(\alpha)}_i} \partial_{z_i} \log\left(\nT^{(\alpha)}\left[u \right] \right) \nd z_i }$


For LO contributions of other cases, they can be transformed to the case of $\bm{b}_L=\bm{b}_R=\bm{-1}$ by some rescaling transformations
\begin{align}
\tilde{u} = u P^\beta \,,\   \tilde{\varphi}_L = \varphi_L/P^\beta        \, , \  \tilde{\varphi}_R = \varphi_R P^\beta   \, . \label{eq:rescaling}
\end{align}
Obviously, $\tilde{u}\tilde{\varphi}_L = u \varphi_L$ and $\tilde{u}^{-1}\tilde{\varphi}_R = u^{-1} \varphi_R$, so this transformation do not change integrals.
Then, one could apply \eref{eq:intnumdlog} and get the formula for the general LO contribution of intersection number
\begin{align}
&\Res_{\bm{\rho}^{(\alpha)}} \nT^{(\alpha)}\left[   \left( \nabla_1^{-1} \cdots \nabla_n^{-1}  \varphi^{(\bm{b}_L)}_L \right)\varphi^{(\bm{b}_R)}_R \right] =
\frac{C_L^{(\bm{b}_L)} C_R^{(\bm{b}_R)}}{\tilde{\bm{\gamma}}^{(\alpha)}} \,, \nn\\
&\tilde{\bm{\gamma}}\equiv \prod_i \tilde{\gamma}_i \,, \quad \tilde{\gamma}_i^{(\alpha)} = \gamma_i^{(\alpha)} - b_{R,i} - 1  \,, \ \ \  \bm{b}_L+\bm{b}_R=\bm{-2} \,  . \label{eq:LO}
\end{align}
where the shifting of $\tilde{\bm{\gamma}}$ comes from the factor $P^\beta$ making $\tilde{\varphi}_R$ having 
$\bm{\tilde{b}}_R=\bm{-1}$.

\subsection{Factorization of poles} \label{sec:2.4}
For a degenerate pole, there are more than one independent integration cycle. We will show its property and factorization transformations correspond to these independent cycles.

\subsubsection{Factorization and contour}

For multivariate residue 
(see \cite{Johansson:2015ava,Larsen:2017aqb}), usually the integration circle is defined by $|P_i(\vec{z})|=r_i$ instead of $|z-z_0|=r$
for the univariate case. However, for the degenerate case, for example, pole $(0,0)$
coming from the intersection of three surfaces  (of denominators) $\{P_1=x_1=0,P_2=x_2=0, P_3=f(x_1,x_2)=0\}$,
the circle defined by taking two of three $P_i$'s is ill-defined. 
To see it, let us consider the integration cycle
\begin{equation}
\oint_{|P_1|=|x_1|=r_1} \nd x_1 \oint_{|P_2|=|x_2|=r_2} \nd x_2 \times \cdots  \, .~~~\label{P-cir}
\end{equation}
Obviously, $x_1=0$ is inside the circle of $x_1=r_1$, and $x_2=0$ is inside the circle of $x_2=r_2$. 
From the implicit function $P_3$, we can solve the variable $x_1$ by the variable $x_2$, which we 
denote as $\bar{f}$, i.e.,
\be x_1 = \bar{f}(x_2) = x_2^{a_1} +\mathcal{O}(x_2^{a_2}) \ee
with $0<a_1<a_2$ for $P_3$  passing through the point $(0,0)$.
The inverse of $\bar{f}$ is given by 
\be x_2 = \bar{f}^{-1}(x_1) = x_1^{1/a_1} +\mathcal{O}(x_1^{\frac{1}{a_1}+\frac{a_2-a_1}{a_1}})\ee
If $\bar{f}(r_2)<r_1$,  $P_3=0$ is in the circle $x_1=r_1$, but meanwhile this implies $\bar{f}^{-1}(r_1)>r_2$, which means $P_3=0$ is not in the circle $x_2=r_2$. We denote this case as $(\{P_2\},\{P_3,P_1\})$. On the contrary, If $\bar{f}(r_2)>r_1$, which implies $\bar{f}^{-1}(r_1)<r_2$, $P_3=0$ is in the cycle of $x_2=r_2$ and not in the circle of $x_1=r_1$. We denote this case as $(\{P_1\},\{P_2,P_3\})$. By selecting different $|P_i|=r_i$ as did in \eref{P-cir}, one can also find another contour corresponds to the combination $(\{P_3\},\{P_1,P_2\})$.
The  analysis tells us  that multivariate residues depend  not only on the location of the pole, but also
 the shape of the cycle enclosing the pole as shown in \cite{Johansson:2015ava, Larsen:2017aqb}. 

As pointed out in \cite{Chestnov:2022xsy}, to deal with
	degenerated cases one can use the method of resolution of singularities, which relates closely to the sector
	decomposition.  To understand the procedure,
	let's show a simple example. Consider $u=z_1^{\beta_1} z_2^{\beta_2} (z_1+z_2)^{\beta_3}\prod_{i=4}^n(\text{C}_i+\mathcal{O}(\bm{z}))^{\beta_i}$. The pole $\bm{p}=(0,0)$ is non-factorized and degenerate, since there are three hypersurfaces 
\begin{equation}
P_1=z_1=0\, , \quad P_2=z_2=0\, , \quad  P_3=z_1+z_2=0
\end{equation}
meet at $(0,0)$ but it is only two dimension (or says 2-variable) problem. 

One could find three transformations:
\begin{align}
&\nT^{(1)}: z_1\to x_1^{(1)}x_2^{(1)} \,,\ z_2\to x_2^{(1)} \, , \nn\\
&\nT^{(2)}: z_1\to x_2^{(2)} \,,\ z_2\to x_1^{(2)}x_2^{(2)} \, , \nn\\
&\nT^{(3)}: z_1\to x_1^{(3)}x_2^{(3)}-x_2^{(3)}  \,,\  z_2\to x_2^{(3)}  \, .~~~\label{3-T}
\end{align}
They are built according to the following logic. For the $n$-variable problem,  first  we choose $n$ $P_i$ as $n$ new variables $x_i$. Then we turn  all the remaining factors $P_j$ which lead to degeneration into the form {  $x_t^a(\text{C}+\mathcal{O}(\bm{z}))$} by a series of transformations {  $x_i\to x_i x_j^b$ with a chosen pair of $(x_i,x_j)$}.
The choice of $b$ is important for the intersection number and we will discuss it later.  For example, we select $P_2$ as $x_2$, $P_3$ as $x_1$. With this choice,  we have the shift transformation
\begin{align}
&{\text t}_1: z_1\to x_1-x_2\,, \ z_2\to x_2\, , \Longrightarrow 
P_1=x_1-x_2\,,\ P_2=x_2\,,\  P_3=x_1\,.
\end{align}
Now we need to turn $P_1$ to the form $x_i^a(\text{C}+\mathcal{O}(\bm{z}))$. There are two different choices. Let us factorize $x_2$ with the second transformation\footnote{Another choice is to factorize $x_1$ with the second transformation $x_2\to x_1^{(3)}x_2^{(3)}\,, \ x_1\to x_1^{(3)}$, which is just $\nT^{(3)}$ with the relabeling of $x_1\leftrightarrow x_2$.}
\begin{align}
&{\text t}_2: x_1\to x_1^{(3)}x_2^{(3)}\,, \ x_2\to x_2^{(3)}\, ,
\end{align}
where the $b=1$. 
Putting these two together, we have $\nT^{(3)}={\text t}_2\circ {\text t}_1$.

Since $x_2$ has been factorized from $P_1$, we combine $P_1$ with $P_2=x_2$ to write  this case as $\{P_3,P_1 P_2\}$\footnote{  Another understanding is to use the "region" concept in \eref{T3-P}. }. As we will show shortly, it just corresponds to the contour $(\{P_3\},\{P_1,P_2\})$. Then, we have
\begin{equation}
\nT^{(1)}: (\{P_1\},\{P_2,P_3\})   \, , \quad \nT^{(2)}: (\{P_2\},\{P_1,P_3\})   \, , \quad   \nT^{(3)}: (\{P_3\},\{P_1,P_2\})  \, . \label{eq:Cofdegenerate}
\end{equation}
From it we can read the hypersurface-powers from $\nT^{(\alpha)}[u]$:
\begin{align}
&\nT^{(1)}[u] =  \left(x_1^{(1)} \right)^{\beta _1} \left(x_2^{(1)}\right)^{\beta
   _1+\beta_2+\beta_3} 
 \left(x_1^{(1)}+1\right)^{\beta _3}  \times \cdots  \, ,\quad
&\gamma^{(1)}_1=\beta_1 \,,\ \gamma^{(1)}_2=\beta_1  +\beta_2+\beta_3 \, , \nn\\
&\nT^{(2)}[u] = \left(x_1^{(2)} \right)^{\beta _2} \left(x_2^{(2)}\right)^{\beta
   _1+\beta _2+\beta _3} \left(x_1^{(2)}+1\right)^{\beta _3}  \times \cdots  \, ,\quad 
& \gamma^{(2)}_1=\beta_2 \,,\ \gamma^{(2)}_2=\beta_1+\beta_2+\beta_3 \, , \nn\\
&\nT^{(3)}[u] = \left(x_1^{(3)}-1\right)^{\beta
   _1}  \left(x_2^{(3)} \right)^{\beta_1 + \beta _2 + \beta_3} 
 \left(x_1^{(3)}\right)^{\beta _3}  \times \cdots  \, ,\quad
&\gamma^{(3)}_1=\beta_3 \,,\ \gamma^{(3)}_2=\beta_1+\beta_2+\beta_3  \, . \label{eq:examfactorize}
\end{align}
Under our regularization scheme, all factors in $\varphi$ are also shown in $u$, so the factorization of $u$ will factorize $\varphi$ automatically.

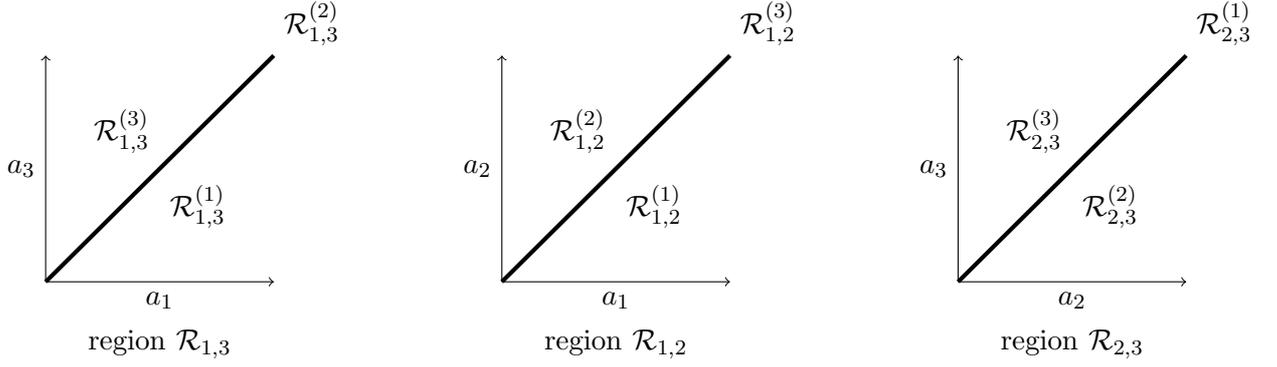
\begin{figure}
	\begin{center}
		\begin{tikzpicture}

			\draw[->] (0,0)--(3,0);
			\draw[->] (0,0)--(0,3);
			\draw[ultra thick] (0,0)--(3,3);
			\node[below] at (1.5, 0){$a_1$};
			\node[left] at (0,1.5){$a_3$};
			\node at (1,2){$\mathcal{R}^{(3)}_{1,3}$};
			\node at (2,1){$\mathcal{R}^{(1)}_{1,3}$};
			\node [above right] at (3,3){$\mathcal{R}^{(2)}_{1,3}$};
			\node[below] at (1.5, -0.5){{\text region} $\mathcal{R}_{1,3}$};
			
			\draw[->] (6,0)--(9,0);
			\draw[->] (6,0)--(6,3);
			\draw[ultra thick] (6,0)--(9,3);
			\node[below] at (7.5, 0){$a_1$};
			\node[left] at (6,1.5){$a_2$};
			\node at (7,2){$\mathcal{R}^{(2)}_{1,2}$};
			\node at (8,1){$\mathcal{R}^{(1)}_{1,2}$};
			\node [above right] at (9,3){$\mathcal{R}^{(3)}_{1,2}$};
			\node[below] at (7.5, -0.5){{\text region} $\mathcal{R}_{1,2}$};
			
			\draw[->] (12,0)--(15,0);
			\draw[->] (12,0)--(12,3);
			\draw[ultra thick] (12,0)--(15,3);
			\node[below] at (13.5, 0){$a_2$};
			\node[left] at (12,1.5){$a_3$};
			\node at (13,2){$\mathcal{R}^{(3)}_{2,3}$};
			\node at (14,1){$\mathcal{R}^{(2)}_{2,3}$};
			\node [above right] at (15,3){$\mathcal{R}^{(1)}_{2,3}$};
			\node[below] at (13.5, -0.5){{\text region} $\mathcal{R}_{2,3}$};
		\end{tikzpicture}
\caption{At the left it is the region of $P_1,P_3$,  the middle  the region of $P_1,P_2$ and   the right  the region of $P_2,P_3$ } \label{region-1}
\end{center}
\end{figure}

Since we are going to take residue around $x_i^{(\alpha)}=0$, 
let us consider  the  limit to $(0,0)$ for each transformation. By the transformation we have 
\bea \nT^{(1)}:&~~~ & P_1=x_1^{(1)} x_2^{(1)},~~~P_2=x_2^{(1)},~~P_3=x_2^{(1)}(x_1^{(1)}+1)\nnn
\nT^{(2)}:&~~~ & P_1=x_2^{(2)},~~~P_2=x_1^{(2)}  x_2^{(2)},~~P_3=x_2^{(2)}(1+x_1^{(2)})\nnn
	\nT^{(3)}:&~~~ & P_1=x_2^{(3)}(x_1^{(3)}-1),~~~P_2=x_2^{(3)},~~P_3=x_1^{(3)} x_2^{(3)}
	~~~\label{T3-P}\eea
The limit going to $(0,0)$ can be described as $x_1\to \lambda^{\beta_1}, x_1\to \lambda^{\beta_2}$ with $\beta_1>0,\beta_2>0$. If we describe the limit of $P_i$ as $P_i\to \lambda^{a_i}$, we will have 
\bea \nT^{(1)}:&~~~ & a_1=\beta_1+\beta_2,~~~~a_2=\beta_2,~~~~~~a_3=\beta_2\nnn
\nT^{(2)}:&~~~ & a_1=\beta_2,~~~~a_2=\beta_1+\beta_2,~~~~~~a_3=\beta_2\nnn
\nT^{(3)}:&~~~ & a_1=\beta_2,~~~~a_2=\beta_2,~~~~~~a_3=\beta_1+\beta_2
~~~\label{T3-P-a}\eea
From it we can read out  the character of  each transformation to be 
\begin{align}
	&\mathcal{R}^{(1)}: \{P_2\to 0,P_3\to 0,\frac{P_1}{P_2}\to 0,\frac{P_1}{P_3}\to 0,\frac{P_3}{P_2}\to 1 \} \, , \nn\\
	&\mathcal{R}^{(2)}:  \{P_1\to 0,P_3\to 0,\frac{P_2}{P_1}\to 0,\frac{P_2}{P_3}\to 0,\frac{P_3}{P_1}\to 1 \}  \, , \nn\\
	&\mathcal{R}^{(3)}: \{P_1\to 0,P_2\to 0,\frac{P_3}{P_1}\to 0,\frac{P_3}{P_2}\to 0,\frac{P_2}{P_1}\to -1 \}  \, ,   \, .~~~\label{reg}
\end{align}
For a better understanding, let us consider the coordinate system of $a_1,a_3$. 
From \eref{T3-P-a} one can  see that  each transformation defines a "region" in this coordinate system  as 
\begin{align}
	\mathcal{R}^{(1)}_{1,3}:\   \{ a_1>0\,,\ a_3>0\,,\ a_1>a_3  \} \, ,\nn\\
	\mathcal{R}^{(2)}_{1,3}:\   \{ a_1>0\,,\ a_3>0\,,\ a_1=a_3  \} \, , \nn\\
	\mathcal{R}^{(3)}_{1,3}:\   \{ a_1>0\,,\ a_3>0\,,\ a_1<a_3  \} \, ,
\end{align}
The combination of these three regions fills the first quadrant as shown in the left picture in Figure~\ref{region-1},
although the measure of $\mathcal{R}^{(2)}_{1,3}$ is zero comparing to other two regions. Same phenomenon also occurs  for $\mathcal{R}_{1,2}$ and $\mathcal{R}_{2,3}$ as shown in  Figure~\ref{region-1}
\begin{align}
	\mathcal{R}^{(1)}_{1,2}:\   \{ a_1>0\,,\ a_2>0\,,\ a_1>a_2  \} \, ,\nn\\
	\mathcal{R}^{(2)}_{1,2}:\   \{ a_1>0\,,\ a_2>0\,,\ a_1<a_2  \} \, , \nn\\
	\mathcal{R}^{(3)}_{1,2}:\   \{ a_1>0\,,\ a_2>0\,,\ a_1=a_2  \} \, ,\nn\\
	\mathcal{R}^{(1)}_{2,3}:\   \{ a_2>0\,,\ a_3>0\,,\ a_2=a_3  \} \, ,\nn\\
	\mathcal{R}^{(2)}_{2,3}:\   \{ a_2>0\,,\ a_3>0\,,\ a_2>a_3  \} \, , \nn\\
	\mathcal{R}^{(3)}_{2,3}:\   \{ a_2>0\,,\ a_3>0\,,\ a_2<a_3  \} \, .
\end{align}

For $\nT^{(1)}$, $P_1$ is much smaller than $P_2$ and $P_3$ means for 
\begin{equation}
\oint_{|P_1|=r_1} \oint_{|P_2|=r_2} \nd P_1 \wedge \nd P_2 \times \cdots  \, .
\end{equation}
we have $r_1<|P_2|$. Hence $P_2=0$ and $P_3=0$ is not in the circle of $|P_1|=r_1$, but in the circle of $|P_2|=r_2$. That's why it does correspond to the contour $(\{P_1\},\{P_2,P_3\})$. One can also consider this problem starting from
\begin{equation}
\oint_{|P_1|=r_1} \oint_{|P_3|=r_2} \nd P_1 \wedge \nd P_3 \times \cdots  \, .
\end{equation}
and get the same conclusion. Similarly, one could know the contour of $\nT^{(2)}$ is $(\{P_2\},\{P_1,P_3\})$ and the contour of $\nT^{(3)}$ is $(\{P_3\},\{P_1,P_2\})$.

Now, as an excise let's compute a multivariate residue  at $(0,0)$ of
\begin{equation}
\varphi=\frac{\left(z_1+z_2\right) R_{1,2}+z_2 R_{1,3}-z_1 R_{2,3}}{z_1 z_2 \left(z_1+z_2\right)}\nd z_1 \nd z_2 \, .
\end{equation}
For this simple case, the denominator of the rational function  can be directly separated to
\begin{equation}
\varphi=\frac{R_{1,2}}{P_1 P_2}\nd P_1 \wedge \nd P_2 + \frac{R_{1,3}}{P_1 P_3}\nd P_1 \wedge \nd P_3+\frac{R_{2,3}}{P_2 P_3}\nd P_2 \wedge \nd P_3
\end{equation}
Using the above analysis for the degenerated case, for the transformation $\nT^{(1)}$ we have 
\bea \varphi|_{\nT^{(1)}}={ (x_1^{(1)}+1) R_{1,2}+ R_{1,3}-x_1^{(1)} R_{2,3}\over x_1^{(1)}x_2^{(1)}(x_1^{(1)}+1) }
dx_1^{(1)}\wedge dx_2^{(1)}\eea
thus
\bea \Res_{(0,0)} \nT^{(1)}[\varphi] = R_{1,2}+R_{1,3}\eea
On the other side, the transformation $\nT^{(1)}$ corresponds to the contour $\{P_1,P_2 P_3\}$. For the analyis of contour, $1/(P_1P_2)$ and $1/(P_1P_3)$ could contribute, thus the result is $R_{1,2}+R_{1,3}$, the same as which we get from transformation $\nT^{(1)}$. 
Similarly
\begin{align}
&R^{(2)}=\Res_{(0,0)} \nT^{(2)}[\varphi] = -R_{1,2}+R_{2,3} \nn\\
&R^{(3)}=\Res_{(0,0)} \nT^{(3)}[\varphi] = -R_{1,3}-R_{2,3} \, .
\end{align}
where the minus comes from $\nd P_i \wedge \nd P_j = -\nd P_j \wedge \nd P_i $. 
 $R^{(1)}+R^{(2)}+R^{(3)}=0$ is due to the three contours are not independent \cite{Larsen:2017aqb}.

Before ending this part, let us mention that there are other  transformations. For example, one could select $P_3$ as $x_1$ and $P_1$ as $x_2$, then, factorize a $x_2$ in the remaining degenerate factor $P_2$. This gives a factorization transformation
\begin{align}
&\nT^{(4)}: z_1\to x_2^{(4)}  \,,\  z_2\to x_1^{(4)}x_2^{(4)}-x_2^{(4)}\, .
\end{align}
which leads to 
\bea P_1=x_2^{(4)},~~~P_2=x_2^{(4)}(x_1^{(4)}-1),~~~~P_3=x_1^{(4)}x_2^{(4)} ~~~\label{T4-1}\eea
and
\begin{align}
&\nT^{(4)}[u] =  \left(x_2^{(4)} \right)^{\beta_1 + \beta _2 + \beta_3} \left(x_1^{(4)}-1\right)^{\beta
   _2} 
 \left(x_1^{(4)}\right)^{\beta _3}  \times \cdots  \, ,
\gamma^{(4)}_1=\beta_3 \,,\ \gamma^{(4)}_2=\beta_1+\beta_2+\beta_3  \,~~~\label{T4-2}
\end{align}
Comparing \eref{T4-1} with $\nT^{(3)}$ in \eref{T3-P}, we see they are same  after noting 
$P_1|_{\nT^{(3)}}\sim x_2^{(3)}$ and $P_2|_{\nT^{(4)}}\sim x_2^{(4)}$. 
Similarly, we have 
$\gamma^{(4)}_1 = \gamma^{(3)}_1$, $\gamma^{(4)}_2 = \gamma^{(3)}_2$ and $\mathcal{R}^{(4)} = \mathcal{R}^{(3)}$ by the analysis of region.
The  contour corresponds to $\nT^{(4)}[u]$ is also $(\{P_3\},\{P_1,P_2\})$, the same as $\nT^{(3)}$. These arguments show it is equivalent to the transformation $\nT^{(3)}$, thus we can neglect it. 

\subsubsection{Region of factorization}

Now let's turn to a more subtle issue about the power $b$ when we do the transformation to factorize the
degenerated pole. The choice of $b$ relates also to the concept of "region" defined by the coordinate system of 
$a_i$, which describes the limit behavior $P_i\sim \lambda^{a_i}$. For the example presented in previous subsubsection, the three factorization transformations $\nT^{(\alpha)}$ in \eref{3-T} are chosen such that
their sum of regions will fill whole region defined by $\mathcal{R}_{i,j}=\{a_i>0,a_j>0\}$ for all pairs of $P_i=0,P_j=0$ as demonstrated in Figure~\ref{region-1}. We claim that only when the factorization transformation
satisfies this  criterion, we will get the right  LO contribution using the formula \eref{eq:intnumdlog}.
Unfortunately, due to our limited mathematical knowledge, the rigorous mathematical reasons for  this phenomenon are unclear to us. But we check our computation to be right via another computation method for multivariate intersection number \cite{Frellesvig:2019uqt}, which does not suffer from the complication of multivariate complex function. Here we illustrate some features of this problem.

This problem arises from that we are not just computing the multivariate residue of a rational function, but also 
the residue of factorization transformation as the denominator in \eref{eq:intnumdlog}.
For a rational function, once it is factorized, further rescale transformation will not change the result. For example
\begin{align}
\Res_{(0,0)}\frac{\nd z_1\wedge \nd z_2}{z_1 z_2} =1~~~~\label{res-1}
\end{align}
A further rescale transformation\footnote{More generally, under the rescale transformation 
$z_1\to t_1^a t_2^b,z_2\to t_1^c t_2^d$, the residue of \eref{res-1} is invariant when $ad-bc=1$.}
\begin{align}
\nT^{(1)}: z_1\to x^{(1)}_1 \left(x^{(1)}_2\right)^2\,, \ z_2\to x^{(1)}_2
\end{align}
will not change the result
\begin{align}
\Res_{(0,0)}\nT^{(1)}\left[\frac{\nd z_1\wedge \nd z_2}{z_1 z_2}\right] =\Res_{(0,0)}\frac{\nd x^{(1)}_1\wedge \nd x^{(1)}_2}{x^{(1)}_1 x^{(1)}_2} = 1 \, .
\end{align}
However,  the formula of LO contribution of intersection number \eref{eq:intnumdlog} contains the 
inverse of residue in the denominator
\begin{align}
	\frac{1}{ \prod_{i} \Res_{\rho^{(\alpha)}_i} \partial_{z_i} \log\left(\nT^{(\alpha)}\left[u \right] \right) \nd z_i} \,
\end{align}
which will change under rescale transformation. 
For example, for  $u=z_1^{\beta_1}z_2^{\beta_2}\prod_{i=3}^n (\text{C}_i+\mathcal{O}(\bm{z}))^{\beta_i}$
which has been completely factorized, formula \eref{eq:intnumdlog} gives 
\begin{align}
\frac{1}{\bm{\gamma}} \equiv \frac{1}{ \prod_{i} \Res_{z_i=0} \partial_{z_i} \log u ~ \nd z_i} =\frac{1}{\beta_1 \beta_2}  \, .~~~\label{rescale-1-1}
\end{align}

Now we consider a rescale transformation
\begin{align}
	\nT^{(1)}: z_1\to x^{(1)}_1 \left(x^{(1)}_2\right)^b \,, \ z_2\to x^{(1)}_2   ~~~\label{rescale-1-2}
\end{align}
such that 
\begin{align}
	&\nT^{(1)}[u] = \left(x^{(1)}_1\right)^{\beta_1}  \left(x^{(1)}_2\right)^{b\beta_1 + \beta_2}\prod_{i=3}^n (\text{C}_i+\mathcal{O}(\bm{z}))^{\beta_i} \nn\\
	&\frac{1}{\bm{\gamma}^{(1)}} =  \frac{1}{ \prod_{i} \Res_{\rho^{(\alpha)}_i} \partial_{z_i} \log\left(\nT^{(1)}\left[u \right] \right) \nd z_i} = \frac{1}{\beta_1(b\beta_1 + \beta_2)} \label{eq:overfactorized1}
\end{align}
Clearly, $1/\bm{\gamma} \neq 1/ \bm{\gamma}^{(1)}$, the further rescale transformation will change the result of the intersection number. The reason is that before the rescale transformation, the region is the complete first quadrant $\mathcal{R}_{1,2}=\{a_1>0,a_2>0\} $. After the rescale transformation, we have 
\bea P_1=z_1= x^{(1)}_1 \left(x^{(1)}_2\right)^b,~~~~P_2=z_2=x^{(1)}_2 \eea
thus the region is given by 
$\mathcal{R}_{1,2}^{(1)}=\{a_1>ba_2>0\} $, which is only part of first quadrant. To get the right result, we must
sum up another rescale transformation, which gives the region  $\mathcal{R}_{1,2}^{(1)}=\{ba_2>a_1>0\} $. One can see that such a rescale transformation is given by 
\begin{align}
	\nT^{(2)}: z_1\to x^{(2)}_1  \,, \ z_2\to x^{(2)}_2\left(x^{(2)}_1\right)^{\frac{1}{b}}   
\end{align}
then
\begin{align}
	&\nT^{(2)}[u] = \left(x^{(2)}_1\right)^{\beta_1+\beta_2/b}  \left(x^{(2)}_2\right)^{\beta_2}\prod_{i=3}^n (\text{C}_i+\mathcal{O}(\bm{z}))^{\beta_i} \nn\\
	&\frac{1}{\bm{\gamma}^{(2)}} =  \frac{1}{ \prod_{i} \Res_{\rho^{(\alpha)}_i} \partial_{z_i} \log\left(\nT^{(2)}\left[u \right] \right) \nd z_i} = \frac{1}{(\beta_1 + \beta_2/b)\beta_2} \, .
\end{align}
It is easy to check that\footnote{More rigirously, there should be  a region $\mathcal{R}_{1,2}^{(1)}=\{ba_2=a_1>0\} $ to completely fill the complete first quadrant. However, the measure of this region is zero, so we can neglect it.}
\begin{align}
	\mathcal{R}_{1,2} = \mathcal{R}^{(1)}_{1,2} +\mathcal{R}^{(2)}_{1,2} \, , \quad \frac{1}{\bm{\gamma}} = \frac{1}{\bm{\gamma}^{(1)}} +\frac{1}{\bm{\gamma}^{(2)}}  \, .   \label{eq:regionrule}
\end{align}

Having the above example, we can come back to the example in the previous subsubsection for the degenerated pole. Let us keep 
factorization transformation of $\nT^{(1)}, \nT^{(2)}$ untouched, but change the factorizaton transformation of $\nT^{(3)}$. Again, in the first step, we do 
\begin{align}
	&{\text t}_1: z_1\to x_1-x_2\,, \ z_2\to x_2\, , \Longrightarrow 
	P_1=x_1-x_2\,,\ P_2=x_2\,,\  P_3=x_1\,.
\end{align}
But in the second step, we do 
\begin{align}
	&{\text t}_2: x_1\to x_1^{(3a)}(x_2^{(3a)})^b\,, ~~~\ x_2\to x_2^{(3a)}\, ,~~~b>1
\end{align}
Overall we have 
\bea  P_1=x_2^{(3a)}(x_1^{(3a)}(x_2^{(3a)})^{b-1}-1),~~~P_2=x_2^{(3a)},~~P_3=x_1^{(3a)} (x_2^{(3a)})^b~~~~\label{T3a-p}\eea
and
\bea & & \nT^{(3a)}[u] = \left(x_1^{(3a)}(x_2^{(3a)})^{b-1}-1\right)^{\beta
	_1}  \left(x_2^{(3a)} \right)^{\beta_1 + \beta _2 +b \beta_3} 
\left(x_1^{(3a)}\right)^{b\beta _3}  \times \cdots  \nnn
&&\gamma^{(3a)}_1=\beta_3 \,,\ ~~~~\gamma^{(3a)}_2=\beta_1+\beta_2+b\beta_3 \eea
It is easy to read out the region by $\nT^{(3a)}$ as 
\bea \mathcal{R}^{(3a)}_{1,2} = \{a_1=a_2>0\},~~~~\mathcal{R}^{(3a)}_{1,3} = \{a_3>ba_1>0\},~~~
\mathcal{R}^{(3a)}_{2,3} = \{a_3>b a_2>0\}\eea
If we check the Figure~\ref{region-1} again, we see that the region $\mathcal{R}_{1,3}$ and $\mathcal{R}_{2,3}$
is not complete. To find the missing region,  we need to consider another transformation
\bea
	&\widetilde{ {\text t}}_2:~~ x_1\to x_1^{(3b)}(x_2^{(3b)})^{1+{1\over b-1}}\,,~~~~ \ x_2\to x_1^{(3b)}(x_2^{(3b)})^{{1\over b-1}}\, 
\eea
Thus we have 
\bea P_1=x_1^{(3b)}(x_2^{(3b)})^{{1\over b-1}}(x_2^{(3b)}-1),~~~~P_2=x_1^{(3b)}(x_2^{(3b)})^{{1\over b-1}},~~~~P_3=x_1^{(3b)}(x_2^{(3b)})^{1+{1\over b-1}}~~~~\label{T3b-p}\eea
and 
\bea & & \nT^{(3b)}[u] = \left((x_2^{(3b)})-1\right)^{\beta
	_1}  \left(x_1^{(3b)} \right)^{\beta_1 + \beta _2 + \beta_3} 
\left(x_2^{(3a)}\right)^{\beta_1+\beta_2+b\beta_3\over b-1}  \times \cdots  \nnn
&&\gamma^{(3b)}_1=\beta_1 + \beta _2 + \beta_3\,,\ ~~~~\gamma^{(3b)}_2={\beta_1+\beta_2+b\beta_3\over b-1} \eea
It is important to notice that both \eref{T3a-p} and \eref{T3b-p} gives the contour $\{ \{P_1,P_2\}, P_3\}$. Furthermore, from \eref{T3b-p} we find the region
\bea \mathcal{R}^{(3b)}_{1,2} = \{a_1=a_2>0\},~~\mathcal{R}^{(3b)}_{1,3} = \{ba_1>a_3>a_1>0\},~
\mathcal{R}^{(3b)}_{2,3} = \{ba_2>a_3>a_2>0\}~~~\eea
In fact, one can check that 
\bea & & {1\over \gamma^{(3a)}}+{1\over \gamma^{(3b)}}={1\over \beta_3(\beta_1+\beta_2+b\beta_3)}+{1\over (\beta_1 + \beta _2 + \beta_3)({\beta_1+\beta_2+b\beta_3\over b-1})}\nnn
& = & {1\over \beta_3(\beta_1+\beta_2+\beta_3)}={1\over \gamma^{(3)}}\eea
after comparing with \eref{eq:examfactorize}.

We emphasize that the region rule is only observations in our practice, not proof. It suggests that If one does not want to lose contributions in this method, one should consider the regions of factorization transformations carefully.


\subsection{Selection rule of CDE---I}

With the above mathematical preparations, we move to the discussion of matrix  $\big( \nds \Omega \big)$ 
defined in \eref{eq:intnumDE}. We will focus on the selection rule for nonzero components of $\big( \nds \Omega \big)$ by two different methods. In this section, we will discuss the computation 
of $\big( \nds \Omega \big)$ without using relative cohomology. Thus for denominators $D_i$ with integer power, we need to add a regulator  $D_i^{\delta_i}$ to $u$ and take the zero limit for the final result. As a consequence,  the CDE selection rule we get in this section will have some redundant terms, which will vanish together with the regulator power $\delta_i$. This problem will be avoided by using relative cohomology in the next
section. The discussion in this section will be the expansion of \cite{Chen:2023kgw}.
From \eref{eq:intnumDE} we see that CDE has two factors:
\begin{align}
	&\big( \nds \Omega \big)_{IK} = \braket{\dot{\varphi}_I|\varphi_J} \big( \eta^{-1} \big)_{JK} \, .
	\label{eq:intnumDE-1}
\end{align}
We will address these two factors one by one.

\subsubsection{Condition of $\braket{\dot{\varphi}_I|\varphi_J}\neq 0$}

Let's consider $\braket{\dot{\varphi}_I|\varphi_J}$ first. By general arguments, to have a nonzero contribution
to intersection number, one must have $b_{\dot{I},i}+b_{J,i} \leq -2$ for series expansion around the pole $p_i$.
For $\varphi_J$ as a $\nd \log$-form, $b_{J,i}\geq -1$, thus only those terms in $\dot{\varphi}_I$ satisfy $b_{\dot{I},i}\leq -1$ lead to non-zero contributions to this intersection number. Furthermore, around
each pole's region $(\alpha)$, the action of $\nds$ operator  decreases only one index $i$ of $b_{I,i}$
in the expansion of $\varphi_I$  by 1. Combining the fact that $\varphi_I$ is also a $\nd \log$-form, we have 
the condition: only when one index $j$ satisfies $b_{I,j}+b_{J,j}\leq -1$  and other indices satisfy  $b_{I,i}+b_{J,i} = -2, i\neq j$, $\braket{\dot{\varphi}_I|\varphi_J}$ could be non-zero.

With the above explanations, now we define two notations. For a pole's region $(\alpha)$, if they have a pair in the expansion which satisfy $b_{I,j}+b_{J,j} = -1$  and $b_{I,i}+b_{J,i} = -2$ for other  $i\neq j$, we say they share a {\bf $(n-1)$-variable Simple pole (($n-1$)-SP)}; if  $b_{I,i}+b_{J,i} = -2$ for all $i$,  we say they share a {\bf $n$-SP}. Using these notations, the above discussions can be summarized as follows:
\begin{align}
&\begin{cases}
\nd \log \text{-form} &: \  b_{I,i}\geq-1 \, , \ b_{J,i}\geq-1 \, , \\
\dot{\varphi}_I &:\  b_{\dot{I},j} = b_{I,j}-1 \, , \ \text{for one $j$} \, , \\
\braket{\dot{\varphi}_I|\varphi_J}\neq 0 &: \ b_{\dot{I},i}+b_{J,i} \leq -2, ~\forall i 
\end{cases} \nn\\
&\Rightarrow \braket{\dot{\varphi}_I|\varphi_J}\neq 0 :\  (-2\leq b_{I,j}+b_{J,j} \leq -1) \  \& \  (b_{I,i}=b_{J,i} = -1, i\neq j)
\end{align}
and 
\begin{align}
\text{($n-1$)-SP} &: \ ( b_{I,j}+b_{J,j} = -1) \, \& \  (b_{I,i}=b_{J,i} = -1, i\neq j) \, , \nn\\
\text{$n$-SP} &:  \ b_{I,i}=b_{J,i} = -1 \, .
\end{align}

Before the computations of the above two cases, let us rewrite 
$\dot{\varphi}_I= \nD \log u \bigwedge_j \nD \log \nW^{(I)}_j$ given in \eref{tot-D} to more explicit form using the expansion \eref{eq:degenerate}
for factorization in the region $(\alpha)$ as  
\begin{align}
 &\nD \log \left( \nT^{(\alpha)} [u] \right) \bigwedge_j \nD \log \nT^{(\alpha)}\left[\nW^{(I)}_j \right] \, ,\nn\\
& \nD \log \left( \nT^{(\alpha)} [u] \right) = \nD \log \bar{u}_\alpha(\bm{x}^{(\alpha)}) + \nD \log \left(\prod_i \left[ x_i^{(\alpha)} - \rho_i^{(\alpha)} \right]^{\gamma_i^{(\alpha)}} \right) \label{eq:TDlog}
\end{align}


\subsubsection{($n-1$)-SP contribution for $\braket{\dot{\varphi}_I|\varphi_J}$}

For $\varphi_I$ and $\varphi_J$ share a ($n-1$)-SP, $\nD \log \left( \nT^{(\alpha)} [u] \right)$ provides the remaining "one pole" via $b_{\dot{I},j}=b_{I,j}-1$.  The result is 
\begin{equation}
    -{\frac{{\gamma}^{(\alpha)}_j }{\bm{{\gamma}}^{(\alpha)}}} \, \nds \int C_{I}^{(\bm{b}_{I} )}C_J^{(\bm{b}_J)} \, d \rho^{(\alpha)}_j \,,
    \label{eq:NMCSPdlog}
\end{equation}
which has been given in \cite{Chen:2023kgw} (include why is it a $\nds \log$), so we are not going to explain these details again, but merely show one typical example: the $C_{I}^{(\bm{b}_{I} )}$ and $C_J^{(\bm{b}_J)}$ in the formula could be 
\begin{align}
&C_{I}^{(b_j=0)} = \left(\partial_{z} \log \tau[z,c_2;c_\pm] \right)\Big|_{z=c_1} = \partial_{c_1} \log \tau[c_1,c_2;c_\pm] \nn\\
&C_J^{(b_j=-1)}=\Res_{z=c_1} \nd \log \tau[z,c_1;c_\pm] =1  \, ,
\end{align}
then 
\begin{align}
\nds \int C_{I}^{(b_j=0)} C_J^{(b_j=-1)} \, d c_1 = \nds \log \tau[c_1,c_2;c_\pm]
\end{align}
Let us remind the reader that there are three cases for the $j$ in $b_{\dot{I},j}=b_{I,j}-1$:
\begin{align}
& b_{I,j}=0 \, , \ \ b_{J,j} = -1 \, , \nn\\
& b_{I,j}=-1 \, , \ \ b_{J,j} = 0 \, ,\nn\\
& b_{I,j}=-\frac{1}{2} \, , \ \ b_{J,j} = -\frac{1}{2} \, .
\end{align}
where the third case could emerge from sqrt-type $\nd \log$. According to the formula of LO contribution \eref{eq:LO}, $\tilde{\gamma}^{(\alpha)}_j$'s in the denominator rely on $b_{I,j}$ and $b_{J,j}$, and are different for these three cases. However, the $\nds$ act on $\int u \varphi_I$ also leads to a external factor $\tilde{\gamma}^{(\alpha)}_j$  at numerator.
For example, to using \eref{eq:intnumdlog}, we need apply the rescaling transformation
\begin{align}
	\tilde{u} = u (z_j-\rho_j^{(\alpha)})^{-b_{I,j}} \, ,
\end{align}
and we have $\tilde{\gamma}^{(\alpha)}_i = \gamma^{(\alpha)}_i$  for $i\neq j$. 
Then, $\dot{\varphi}$ take the form as follow:
\begin{align}
	&~~~~\nds (z_j-\rho_j^{(\alpha)})^{\tilde{\gamma}^{(\alpha)}_j} \times \prod_{i\neq j}(z_i-\rho_i^{(\alpha)})^{\gamma^{(\alpha)}_i-1} f \nn\\
	&= -\tilde{\gamma}^{(\alpha)}_j \nds \rho_j^{(\alpha)} \times \prod_{i}(z_i-\rho_i^{(\alpha)})^{\tilde{\gamma}^{(\alpha)}_i-1} f + \cdots \,
\end{align}
where the $\cdots$ part does not contribute to the intersection number. 
The $\tilde{\gamma}^{(\alpha)}_j$ in the above equation will cancel the $\tilde{\gamma}^{(\alpha)}_j$ in $\bm{\tilde{\gamma}}^{(\alpha)}$ in the equation \eref{eq:LO}:
\begin{equation}
	\frac{{\tilde{\gamma}}^{(\alpha)}_j }{\bm{{\tilde{\gamma}}}^{(\alpha)}}= \frac{1}{\prod_{i\neq j} \gamma^{(\alpha)}_i }  = \frac{\gamma^{(\alpha)}_j }{\bm{{\gamma}}^{(\alpha)}} \, .
\end{equation}
Hence, we have the coefficient in \eref{eq:NMCSPdlog}.


\subsubsection{$n$-SP contribution for $\braket{\dot{\varphi}_I|\varphi_J}$}

For $\varphi_I$ and $\varphi_J$ sharing a $n$-SP,  at first glance,  the second term  $\nD \log \left(\prod_i \left[ x_i^{(\alpha)} - \rho_i^{(\alpha)} \right]^{\gamma_i^{(\alpha)}} \right)$ in $\nD \log \left( \nT^{(\alpha)} [u] \right)$ (see \eref{eq:TDlog})  seems to lead to a pair of expanded terms that satisfy $b_{L,j}+b_{R,j}=-3$ for one $j$, and $b_{L,i}+b_{R,i}=-2$ for other $i\neq j$. 
This could give a NLO contribution (as defined in \cite{Chen:2023kgw}) to the intersection number. However, as we will show shortly, all these potential contributions are canceled here. For rational-type $\nd \log$-form ,
\begin{equation}
\nD \log \nT^{(\alpha)}\left[\nW^{(I)}_j \right] = \nD \log \left(x_j^{(\alpha)}-\rho_j^{(\alpha)}\right)
\end{equation} 
expanding $D=\nds+\nd$ where  $\nds$ is differentiation with respect to $\rho_j^{(\alpha)}$ and $\nd_j$  differentiation with respect to $x_j^{(\alpha)}$, we have
\begin{align}
& ~\left[ \nd_{j} \log \left(\left( x_j^{(\alpha)} - \rho_j^{(\alpha)} \right)^{\gamma_j^{(\alpha)}} \right) 
\wedge \nds \log \left( x_j^{(\alpha)} - \rho_j^{(\alpha)} \right)  \right. \nn\\
&\left. ~ + \nds \log \left( \left( x_j^{(\alpha)} - \rho_j^{(\alpha)} \right)^{\gamma_j^{(\alpha)}} \right) 
\wedge \nd_j \log \left( x_j^{(\alpha)} - \rho_j^{(\alpha)} \right)  \right]   \bigwedge_{i\neq j}\nD \log \nT^{(\alpha)}\left[\nW^{(I)}_i \right]   \nn\\
= & \gamma_j^{(\alpha)} \left[ \nd_{j} \log \left( x_j^{(\alpha)} - \rho_j^{(\alpha)} \right)   
\wedge \nds \log \left( x_j^{(\alpha)} - \rho_j^{(\alpha)} \right) \right. \nn\\
&\left. ~~~~ + \nds \log \left( x_j^{(\alpha)} - \rho_j^{(\alpha)} \right)\wedge \nd_j \log \left( x_j^{(\alpha)} - \rho_j^{(\alpha)} \right)  \right]   \bigwedge_{i\neq j}\nD \log \nT^{(\alpha)}\left[\nW^{(I)}_i \right]   \nn\\
=&0 \, .
\end{align}
Similarly, for sqrt-type $\nd \log$-form 
\begin{equation}
\nD \log \nT^{(\alpha)}\left[\nW^{(I)}_j \right] = \nD \log \tau\left[x_j^{(\alpha)} ,\rho_j^{(\alpha)};c_{\pm} \right]
\end{equation} 
we have
\begin{align}
& ~\left[ \nd_{j} \log \left(\left( x_j^{(\alpha)} - \rho_j^{(\alpha)} \right)^{\gamma_j^{(\alpha)}} \right) \wedge \nds \log \tau\left[x_j^{(\alpha)} ,\rho_j^{(\alpha)};c_{\pm} \right]  \right. \nn\\
&\left. +  \nds \log \left(\left( x_j^{(\alpha)} - \rho_j^{(\alpha)} \right)^{\gamma_j^{(\alpha)}} \right) \wedge \nd_j \log \tau\left[x_j^{(\alpha)} ,\rho_j^{(\alpha)};c_{\pm} \right]   \right] 
\bigwedge_{i\neq j}\nD \log \nT^{(\alpha)}\left[\nW^{(I)}_i \right]   \nn\\
=& \left[ \nd_{j} \log \left( x_j^{(\alpha)} - \rho_j^{(\alpha)} \right)\wedge \nds \log \left( x_j^{(\alpha)} - \rho_j^{(\alpha)} \right)  + \nds \log \left( x_j^{(\alpha)} - \rho_j^{(\alpha)} \right)\wedge \nd_j \log \left( x_j^{(\alpha)} - \rho_j^{(\alpha)} \right)  \right] \nn\\
&  \times \frac{\gamma_j^{(\alpha)} \sqrt{\left(\rho_j^{(\alpha)}-c_+\right) \left(\rho_j^{(\alpha)}-c_-\right)}   }{ \sqrt{\left(x_j^{(\alpha)}-c_+ \right) \left(x_j^{(\alpha)}-c_- \right)} }  
\bigwedge_{i\neq j}\nD \log \nT^{(\alpha)}\left[\nW^{(I)}_i \right]   \nn\\
=&0
\end{align}

After showing the second term giving non contribution, using the result \eref{eq:LO} the first term  gives the
result
\be
\frac{C_I^{(\bm{-1})} C_J^{(\bm{-1})}}{\bm{\gamma}^{(\alpha)}} \, \nds \log \Big( \bar{u}_\alpha(\bm{\rho}^{(\alpha)}) \Big) \,.  \label{eq:CSPdlog}
\end{equation}
It is worth pointing out that 
in \cite{Chen:2023kgw}, one needs a formula of the NLO contribution of intersection number for this result. Here we avoid using NLO formula because we have transformed the related term to LO case via IBP in  \eref{eq:Dlogvarphi} 
(from the second line to the third line) at the beginning.


\subsubsection{Condition of $\big( \eta^{-1} \big)_{JK}\neq 0$}
For $\big( \eta^{-1} \big)_{JK}$  in CDE, since $\eta^{-1}={1\over |\eta|}\eta^*$ where $\eta^*$ is the adjugate matrix, the element could be written as 
\begin{equation}
\big( \eta^{-1} \big)_{JK} = (-1)^{J+K} \frac{\eta^{(KJ)}}{|\eta|}
\end{equation}
where $\eta^{(KJ)}$ is the minor of the element $\eta_{JK}$ in the matrix  $\eta$. The minor in the Laplace expansion of the determinant is the sum of the form
\begin{equation}
	(-1)^a \eta_{J i_1}\eta_{i_1 i_2}\cdots \eta_{i_{v-1} i_v}\eta_{i_v K} \prod_{k}\eta_{j_k j_k} \, .~~\label{eta-form}
\end{equation}
For $\eta^{(KJ)}\neq 0$, at least one term of the form \eref{eta-form} should be nonzero, which 
is equivalent to the statement that every element $\eta_{ij}$ in \eref{eta-form} should be zero. For the dlog-form
$\eta_{JK} = \braket{\varphi_J|\varphi_K}$ could be non-zero only when  $\varphi_J$ and  $\varphi_K$ share at least one $n$-SP. Since 
 $\varphi_{j_k}$ must exhibit $n$-SP with itself, all $\eta_{j_k j_k}$ should naturally be non-zero. Therefore, we only need to ask  the existence of at least one non-zero chain $\eta_{J i_1}\eta_{i_1 i_2}\cdots \eta_{i_{v-1} i_v}\eta_{i_v K}$, which implies that $\varphi_J$ shares $n$-SP with $\varphi_{i_1}$, $\varphi_{i_1}$ shares $n$-SP with $\varphi_{i_2}$, and so on. With this picture, the notation of 
{\bf  $n$-SP chain} has been defined  in \cite{Chen:2023kgw}: 
\begin{itemize}
\item If $\varphi_I$ and $\varphi_J$ share an $n$-SP, we say they are $n$-SP related, and denote it as $\varphi_I \sim \varphi_J$. 
\item The $n$-SP chain is the collection of $\varphi_I$'s, such that for arbitrary pair of $\varphi_a, \varphi_b$
there exists an ordered list $\{\varphi_a, \varphi_{i_1},...,\varphi_{i_k}, \varphi_b \}$ such that
$\varphi_a\sim \varphi_{i_1}\sim ...\sim \varphi_{i_k}\sim\varphi_b $ where every $\varphi$ belongs to the chain.  If $\varphi_a, \varphi_b$ belong to a $n$-SP chain, we denote it as $\varphi_a\sim\sim\varphi_b$. 

\end{itemize}
Using this notation, we can simply say that the  condition of $\big( \eta^{-1} \big)_{JK}\neq 0$  is  $\varphi_J$ and $\varphi_K$ belongs to an $n$-SP chain.
For example, for the case with seven master integrals, whose indices are denoted as $\{J,K,1,2,3,4,5\}$,  supposing $\varphi_J\sim\varphi_1\sim\varphi_2\sim\varphi_K$, we have  $\varphi_J\sim\sim\varphi_K$ and 
\begin{equation}
(-1)^{J+K}|\eta^{(KJ)}| = \left|
\begin{array}{cccccc}
 * & \eta _{J1} & * & * & * & *
   \\
 * & * & \eta _{12} & * & * & * \\
 \eta _{2K} & * & * & * & * & * \\
 * & * & * & \eta _{33} & * & * \\
 * & * & * & * & \eta _{44} & * \\
 * & * & * & * & * & \eta _{55} \\
\end{array}
\right|
\end{equation}
could be non-zero due to $\eta _{\text{J}1}\eta _{12}\eta _{2\text{K}}\eta _{33}\eta _{44}\eta _{55}$ could be non-zero.


\subsubsection{Conclusion}

Having discussed two factors in \eref{eq:intnumDE}, now we can state the CDE selection rules:
\begin{itemize}
\item $\big( \nds \Omega \big)_{IK} = \braket{\dot{\varphi}_I|\varphi_J} \big( \eta^{-1} \big)_{JK}$ could be non-zero only when there exist a $\varphi_J$ share ($n-1$)-SP or $n$-SP with $\varphi_I$, and $\varphi_J\sim\sim\varphi_K$.
\item $\big( \nds \Omega \big)_{IK}$ could be determined via merely LO contribution formula of intersection number. For $\varphi_I$ and $\varphi_J$ share a  ($n-1$)-SP, it contributes
\begin{equation}
    -{\frac{{\gamma}^{(\alpha)}_j }{\bm{{\gamma}}^{(\alpha)}}} \, \nds \int C_I^{(\bm{b}_I)}C_J^{(\bm{b}_J)} \, \nds \rho^{(\alpha)}_j \,
    \label{eq:NMCSPdlog1}
\end{equation}
to $\braket{\dot{\varphi}_I|\varphi_J}$. For $\varphi_I$ and $\varphi_J$ share a  $n$-SP, it contributes
\begin{equation}
\frac{C_I^{(\bm{-1})} C_J^{(\bm{-1})}}{\bm{\gamma}^{(\alpha)}} \, \nds \log \Big( \bar{u}_\alpha(\bm{\rho}^{(\alpha)}) \Big) \,.  \label{eq:CSPdlog1}
\end{equation}
to $\braket{\dot{\varphi}_I|\varphi_J}$ with constant $C_I^{(\bm{-1})} C_J^{(\bm{-1})}$. These two formulas give all symbol letters. For $\eta_{JK}=\braket{\varphi_J|\varphi_K}$, each shared $n$-SP gives
\begin{equation}
 \frac{C_J^{(\bm{-1})} C_K^{(\bm{-1})}}{\bm{\gamma}^{(\alpha)}}  \label{eq:inveta}
\end{equation}
with constant $C_J^{(\bm{-1})} C_K^{(\bm{-1})}$.

\item Canonical-form differential equations (CDE) emerge naturally. 
To see it, let's assign the original hypersurface-powers $\beta_i$ in $u=\prod_i \left[P_i(\bm{z})\right]^{\beta_i}$ to  haave the transcendental weight-($-1$). For our application in Feynman integrals, $\beta_i\sim a  \ep$ or $\beta_i \sim \delta_k$ with constant $a$ (typically, integer or half-integer). This means that $\ep$ and $\delta_i$ have the transcendental weight-($-1$), so all hypersurface-powers $\gamma^{(\alpha)}_i$ for each factorization are also weight-($-1$)
\begin{equation}
\mathcal{T}(\beta_i) = \mathcal{T}(\ep) = \mathcal{T}(\delta_k) = \mathcal{T}(\gamma^{(\alpha)}_i) = -1\, .
\end{equation}
Because it \eref{eq:NMCSPdlog1}, \eref{eq:CSPdlog1} as well as $\braket{\dot{\varphi}_I|\varphi_J}$ are weight-($n$-$1$) coefficient times $\nds \log$-form. Similarly  $\big( \eta^{-1} \big)_{JK}$
is weight-($-n$) coefficient. Combining them, we have $\big( \nds \Omega \big)_{IK}$ is weight-($-1$) coefficient times $\nds \log$-form.  After taking regulators $\delta_i$ to  zero, the weight-($-1$) coefficient in $\big( \nds \Omega \big)_{IK}$ could only be proportional to $\ep$. Hence we have 
\begin{equation}
\big( \nds \Omega \big)_{IK} = \ep \sum_i \text{C}^{(i)}_{IK} ~ \nds \log \text{W}^{(i)}(\bm{s}) \label{eq:CDE}
\end{equation}
where $\text{W}^{(i)}(\bm{s})$ are symbol letters and $\text{C}^{(i)}_{IK}$ are the corresponding  constant coefficient matrix. This is nothing, but the so-called canonical form or $\ep$-form.
\end{itemize}
Since after factorization, the n-variable multivariate residues are reduced to n univariate residues, and only the ($n_0\geq$  $n-1$)-SP shared by $\varphi_I$ and $\varphi_J$  could contribute, naturally one can image that people could take residue of $(n-1)$-SP first. This gives an overall factor  
\begin{equation}
{\frac{{\gamma}^{(\alpha)}_j }{\bm{{\gamma}}^{(\alpha)}}}
\end{equation}
and left an univariate problem, whose u part (denoted as $u'$ here) is
\begin{equation}
u'=\bar{u}_\alpha(\rho_1^{(\alpha)},\cdots,\rho_{j-1}^{(\alpha)}, x_j^{(\alpha)},\rho_{j+1}^{(\alpha)},\cdots,\rho_{n}^{(\alpha)} ) \left( x_j^{(\alpha)} - \rho_j^{(\alpha)} \right)^{\gamma_j^{(\alpha)}}
\end{equation}
All univariate u-part could be reformed as 
\begin{equation}
u' =P_0^{\beta_0} \prod_i (z-c_i)^{\beta_i} \,.
\end{equation}
Then, $\nd \log$ basis and CDE of  all  univariate cases without elliptic integrals could be systematically discussed based on this form and of course give the same answer as \eref{eq:NMCSPdlog1} and \eref{eq:CSPdlog1}. Since univariate rationalization is much easier, one could also choose to transform these univariate cases without elliptic integrals into
\begin{equation}
u'=(P_0')^{a_0 \ep} \prod_i (z'-c_i')^{a_i \ep} \,.
\end{equation}
Then all symbol letters are the distance between these univariate poles $c_i'-c_j'$ and pure parameter factor $P_0'$.
Details of this univariate discussion have been given in \cite{Chen:2023kgw}.


\section{Selection rule of CDE via relative cohomology} \label{sec:RC}

\subsection{Dual form in twisted relative cohomology}
In the last section, we have introduced how to compute the intersection number between forms in twisted cohomology. In this section, we will 
present simple computational rules for intersection numbers between twisted cohomology and twisted relative cohomology. 
Recall the integrals take the form
\begin{align}
&I[u,\varphi] \equiv \int  u \, \varphi_L  \,,   \nn \\
&\varphi \equiv \hat{\varphi}(\bm{z}) \bigwedge_j \nd z_j =\frac{Q(\bm{z})}{\left(\prod_k D_k^{a_k} \right) \left(\prod_i P_i^{b_i} \right)} \bigwedge_j \nd z_j \, , \nn\\
&u=\prod_i \left[P_i(\bm{z})\right]^{\beta_i} \,,
    \quad \quad  a_k, b_j\in \mathbb{N}  \, ,
    \label{eq:baikov1}
\end{align}
where regulators have not been introduced for factors $ D_k$.  
In Feynman integrals, the  denominators $D_i$ with integer power are propagators and determine the “sector" to which the integral belongs to. Here we borrow the concept of "sector" from the Feynman integral by considering  denominators in $\varphi_I$ having negative integer power in $u\varphi_I$\footnote{The $P_i$ in $\varphi_I$  will have contribution from $u$, so the total power is not integer and we should consider them as the "propagator". }
as the "propagator" and defining the sector by the list $\hat{I}$ selecting from the set of propagators. Furthermore, we denote the set of hypersurfaces relating to the sector $\hat{I}$ as  
\begin{equation}
\mathcal{B}_{\hat{I}} = \{D_{I_1}=0,D_{I_2}=0,\cdots,D_{I_n}=0  \} \, ~~~\label{B-cut}
\end{equation}
Selecting the dual space of dual form $\ket{\varphi}$ to be the space of twisted  relative cohomology just means selecting  dual forms living on the maximal cut of each sector. To distinguish, we denote these dual forms living on its maximal cut $(\hat{R})$ as $\Delta_{\hat{R}}\varphi_R$. In practice, we do not need to know the exact meaning of "live on the cuts", and all we need to know is the intersection number of $\varphi_L$ and $\Delta_{\hat{R}}\varphi_R$ obey a simple rule: just applying the maximal cut of the sector of $\varphi_R$ to both sides, then applying the intersection number between the cut forms as we did in the previous section, i.e.,
\begin{align}
&\braket{\varphi_L|\Delta_{\hat{R}}\varphi_R}= \braket{\varphi_{L;\hat{R}}  |\varphi_{R;\hat{R}}} = \sum_{\bm{p}_{\hat{R}}}\Res_{\bm{p}_{\hat{R}}=0} \psi_{L;\hat{R}} \varphi_{R;\hat{R}}   \, , \quad \nabla_{1;\hat{R}} \cdots \nabla_{n;\hat{R}} \psi_{L;\hat{R}} = \varphi_{L;\hat{R}} \nn\\
& \varphi_{L;\hat{R}} = \Res_{\mathcal{B}_{\hat{R}}}  \left( \frac{u\varphi_L}{u_{\hat{R}}}\right)  \, , \quad  \varphi_{R;\hat{R}} = \Res_{\mathcal{B}_{\hat{R}}}  \left( \frac{\varphi_R u_{\hat{R}}}{u} \right)      \, , \quad  u_{\hat{R}}=u|_{\mathcal{B}_{\hat{R}}}~~~\label{3.3}
\end{align}
where $\nabla_{i;\hat{R}}$  is defined via the cut integral family corresponding to $u_{\hat{R}}$.  $\bm{p}_{\hat{R}}$ are again isolated intersection points (see \eref{eq:intnum}) of hypersurfaces
containing  $\mathcal{B}_{\hat{R}}$. The definition of $\varphi_{L;\hat{R}}$ in \eref{3.3} has the property
$\Res_{\mathcal{B}_{L;\hat{R}}}  u \varphi = u_{\hat{R}} ~ \varphi_{L;\hat{R}}$, and the form in \eref{3.3}
makes $\varphi_{L;\hat{R}}$ manifest as the single-valued function.

In relative cohomology, the concept of "subsector" is the same as Feynman integrals, i.e., if $\mathcal{B}_{\hat{I}}  \subseteq \mathcal{B}_{\hat{J}} $, we say $\mathcal{B}_{\hat{I}}$ is a subsector of $\mathcal{B}_{\hat{J}}$. Obviously, only when the sector of $\varphi_R$ is a subsector of the sector of $\varphi_L$, the intersection number could be non-zero.

\subsection{Improved CDE selection rule}

Now we consider the $\nds \Omega$ in CDE using the dual basis in the relative cohomology
\begin{align}
    &\big( \nds \Omega \big)_{IK} = \braket{\dot{\varphi}_{I;\hat{J}}|\varphi_{J;\hat{J}}} \big( \eta^{-1} \big)_{JK} \, ,\nn\\
    & \eta_{IJ}\big( \eta^{-1} \big)_{JK} = \delta_{IK}   \, , \ \         \eta_{IJ} = \braket{\varphi_{I;\hat{J}}|\varphi_{J;\hat{J}}} .
    \label{eq:intnumDEc}
\end{align}
For $\braket{\dot{\varphi}_{I;\hat{J}}|\varphi_{J;\hat{J}}}$, let us denote the number of remaining integration variables of sector $\mathcal{B}_{\hat{J}}$ as $n_{\hat{J}}$. With the knowledge discussed in the previous section, we see immediately that  $\braket{\dot{\varphi}_{I;\hat{J}}|\varphi_{J;\hat{J}}}$ could be non-zero only when $\dot{\varphi}_{I;\hat{J}}$ and $\varphi_{J;\hat{J}}$ share $n_{\hat{J}}$-SP or ($n_{\hat{J}}-1$)-SP.

For $\big( \eta^{-1} \big)_{JK}$, thing is a little bit different since
$\braket{\varphi_{I;\hat{J}}|\varphi_{J;\hat{J}}}\neq \braket{\varphi_{J;\hat{I}}|\varphi_{I;\hat{I}}}$. 
To count the anti-symmetry, we need to slightly modify the concept of $n$-SP chain to a new concept of {\bf cut-$n$-SP} chain, i.e.,
\begin{itemize}
\item If ${\varphi}_{I;\hat{J}}$ and $\varphi_{J;\hat{J}}$ share $n_{\hat{J}}$-SP,  we say $\varphi_I$ is cut-$n$-SP related to $\varphi_J$, and denoted as $\varphi_I \rightarrow \varphi_J$ or $\varphi_J \leftarrow \varphi_I$. Now, it is an orient relation, i.e., $\varphi_I \rightarrow \varphi_J$ does not imply $\varphi_J \rightarrow \varphi_I$. 

\item If $\varphi_I \rightarrow \varphi_J$, we also say that $\varphi_I$ is linked to $\varphi_J$ via a cut-$n$-SP chain. If $\varphi_I \rightarrow \varphi_J$ and $\varphi_J \rightarrow \varphi_K$, then we say $\varphi_I$ is linked to $\varphi_K$ via the cut-$n$-SP chain $\varphi_I \rightarrow \varphi_J \rightarrow \varphi_K$, we denote it as $\varphi_I \rightarrow\rightarrow \varphi_J$. Similar understanding for more forms $\varphi$.  
\end{itemize}
With above definition the condition of $\big( \eta^{-1} \big)_{JK}$ could be non-zero is  $\varphi_J \rightarrow\rightarrow \varphi_K$. Obviously, if  $\mathcal{B}_{\hat{I}}  \subset \mathcal{B}_{\hat{J}}$, $\varphi_I \rightarrow \rightarrow \varphi_J$ could not be true.

The contributions for $\braket{\dot{\varphi}_{I;\hat{J}}|\varphi_{J;\hat{J}}}$  are still calculated via \eref{eq:NMCSPdlog1} and \eref{eq:CSPdlog1}. However, every element in these expressions should be replaced by the cut one, corresponding to the maximal cut  $\mathcal{B}_{\hat{J}}$. One difference the cut leads to is the transcendent weight of coefficient $\gamma^{(\alpha)}_j/\bm{\gamma}^{(\alpha)}$,  which changes from $n-1$ to   $n_{\hat{J}}-1$. Similarly, contributions for $\left(\eta^{-1}\right)_{JK}$ is calculated via \eref{eq:inveta}, but the weight of coefficient becomes $-n_{\hat{J}}$. One point that needs to be emphasized is that
since we avoid regulators at the beginning, only $a \ep$ appears in $\beta_i$ and hypersurface-powers.  Weight-n coefficient could only be proportional  to $\ep^{-n}$.

With the above discussions,  we have CDE selection rules (improved and exact version):
\begin{itemize}
\item $\big( \nds \Omega \big)_{IK} = \braket{\dot{\varphi}_{I;\hat{J}}|\varphi_{J;\hat{J}}} \big( \eta^{-1} \big)_{JK}$ could be non-zero only when there exist a $\varphi_J$ that  $\dot{\varphi}_{I;\hat{J}}$ and $\varphi_{J;\hat{J}}$ share $n_{\hat{J}}$-SP or ($n_{\hat{J}}-1$)-SP, and $\varphi_J \rightarrow \rightarrow \varphi_K$.

\item $\big( \nds \Omega \big)_{IK}$ could be determined via merely LO contribution formula of intersection number. It gives cut version of \eref{eq:NMCSPdlog1} and \eref{eq:CSPdlog1} for $n_{\hat{J}}$-SP or ($n_{\hat{J}}-1$)-SP contribution in $\braket{\dot{\varphi}_{I;\hat{J}}|\varphi_{J;\hat{J}}}$ and cut vertion of \eref{eq:inveta} for $\left(\eta^{-1}\right)_{JK}$.
 
\item  $\braket{\dot{\varphi}_{I;\hat{J}}|\varphi_{J;\hat{J}}} \big( \eta^{-1} \big)_{JK}$ is  weight-($-1$) coefficient, which could only be propotional to $\ep$, times  $\nds \log$-form. Thus canonical differential equation emerges.
\end{itemize}

We know that the differential of a Feynman integral belonging to a subsector will not rely on a higher sector, so
\begin{align}
\text{If } \mathcal{B}_{\hat I} \subset \mathcal{B}_{\hat K}\, , \ \text{then } \nds \Omega_{IK} = 0  \, .  \label{eq:relyonsub}
\end{align}
Unfortunately, with regulators the structure of the sector has been broken, i.e., there is also no explicit reason to tell us if $\mathcal{B}_{\hat I} \subset \mathcal{B}_{\hat K}$. Thus the statement \eref{eq:relyonsub} is not true and  $\nds \Omega_{IK}$  vanishes when and only when taking regulators to  zero after the computation of intersection number. This is one of the manifestations of redundancy when using the regulator method.
However, with dual form in relative cohomology,  $\braket{\dot{\varphi}_{I;\hat{J}}|\varphi_{J;\hat{J}}}$ could be non-zero only when $\mathcal{B}_{\hat{J}} \subseteq \mathcal{B}_{\hat{I}}$. Furthermore, $(\eta^{-1})_{JK}$ could be non-zero only when $\mathcal{B}_{\hat{K}} \subseteq \mathcal{B}_{\hat{J}}$. Therefore, $\big( \nds \Omega \big)_{IK}$ could be non-zero only when $\mathcal{B}_{\hat{K}} \subseteq \mathcal{B}_{\hat{I}}$ and \eref{eq:relyonsub} holds true as well.


\section{Univariate example: $u=z^\delta (z-c_1)^\ep(z-c_2)^\ep$} \label{sec:eg1}

In this section, we consider the case $u=z^\delta (z-c_1)^\ep(z-c_2)^\ep$, where $\delta$ is a regulator. This structure appears in many cases of Feynman integral. For example, all n-point one-loop integrals with their $n-1$ propagator cut will give such a u in their Baikov representation. We will compute its CDE via both methods, with and without regulators. 

For the given $u$, the related hypersurfaces are
\begin{align}
\mathcal{B}=\{z=0,z-c_1=0,z-c_2=0,z=\infty\}
\end{align}
Its $\nd \log$ basis can be easily constructed as
\begin{align}
\varphi_1 = \frac{\nd z}{z} =\nd \log z \, ,  \quad \varphi_2 = \frac{\nd z}{z-c_1} - \frac{\nd z}{z-c_2}  = \nd \log \left( \frac{z-c_1}{z-c_2}\right)  \, .
\end{align}

\subsection{With regulator}
With a regulator, we have
\begin{align}
\omega=\nd \log u = \left( \frac{\delta}{z} + \frac{\ep}{z-c_1}+ \frac{\ep}{z-c_2} \right) \nd z \, .
\end{align}
Poles of $\omega$ are
\begin{align}
\{\bm{p}\}=\{0,c_1,c_2,\infty\} \, .
\end{align}
and the hypersurface powers corresponding to each pole are
\begin{align}
\gamma_1=\delta\, , \ \ \gamma_2=\ep\, , \ \ \gamma_3=\ep\, , \ \ \gamma_4=-2\ep-\delta \, .
\end{align}
Now we compute $\eta_{IJ}=\braket{\varphi_I|\varphi_J}$ using the result \eref{eq:intnumexpand} and \eref{eq:LO}. 
For $\eta_{11}$,  $\varphi_1$ shares 1-SP with itself at the points $0$ and $\infty$. For $\eta_{12}$,
$\varphi_1$ do not share any pole with $\varphi_2$. For  $\varphi_2$, it  shares 1-SP with itself at the points $c_1$ and $c_2$.  Thus we have  we have 
\begin{align}
&\eta=\braket{\varphi_I|\varphi_J}=\left(
\begin{array}{cc}
 \frac{1}{\gamma_1}+\frac{1}{\gamma_4} & 0 \\
 0 &  \frac{1}{\gamma_2}+\frac{1}{\gamma_3}  \\
\end{array}
\right)=\left(
\begin{array}{cc}
 \frac{1}{\delta }-\frac{1}{\delta +2 \epsilon } & 0 \\
 0 & \frac{2}{\epsilon } \\
\end{array}
\right)  \, , \nn\\
&\eta^{-1}=\left(
\begin{array}{cc}
 \delta(\delta+2\ep)/(2 \epsilon)  & 0 \\
 0 & \epsilon/2 \\
\end{array}
\right)  \, .
\end{align}
Let's turn to consider $\braket{\dot{\varphi}_I|\varphi_J}$. Using \eref{eq:Dlogvarphi} we have $\dot{\varphi}_I$
\begin{align}
&\dot{\varphi}_1 = \nD \left(\log \bar{u}_1(z) + \delta \log z \right) \wedge \nD \log z \nn\\
& \quad = \nD \log \bar{u}_1(z) \wedge \nD \log z= \nd \log \bar{u}_1(z) \wedge \nds \log z+\nds \log \bar{u}_1(z) \wedge \nd \log z \, , \nn\\
&\dot{\varphi}_2 =  \nD \log \bar{u}_2(z) \wedge \nD \log (z-c_1) - \nD \log \bar{u}_3(z) \wedge \nD \log (z-c_2)   \nn\\
&\quad = \delta ~ \nD \log z \wedge \varphi_2 + 2 \ep ~ \nD \log (z-c_2) \wedge \nD \log (z-c_1)  \, , ~~~\label{4.7}\end{align}
where $\bar{u}_i(z)$ is defined in \eref{eq:degenerate-1} and for current example they are 
\bea 
\bar{u}_1(z) = (z-c_1)^\ep(z-c_2)^\ep \, ,  
\bar{u}_2(z) = z^\delta(z-c_2)^\ep \, ,
\bar{u}_3(z) = z^\delta(z-c_1)^\ep \, . 
\eea
As a pedagogical example, we show the computation details of  $\braket{\dot{\varphi}_1|\varphi_1}$ here. 
First as the d-log form, $\varphi_1$ shares 1-SP with itself only, i.e., there is no contribution from 
0-SP. For 
pole $z=0$, the first term of $\dot{\varphi}_1$ in \eref{4.7}  gives zero contribution, while  the second term gives the contribution
\begin{align}
\frac{1}{\delta} \nds \log \left((z-c_1)^\ep(z-c_2)^\ep|_{z=0} \right) = \frac{\ep}{\delta} \nds \log(c_1c_2)  \,,
\end{align}
For the pole $z=\infty$ the second term gives the contribution
\begin{align}
\frac{1}{-\delta-2\ep} \nds \log \left((1-c_1 t)^\ep(1-c_2 t)^\ep|_{t=0} \right) = \frac{\ep}{-\delta-2\ep} \nds \log 1 = 0 \,,
\end{align}
where $t=1/z$. 
For $\braket{\dot{\varphi}_1|\varphi_2}$, $\varphi_1$ shares the 0-SP with $\varphi_2$ at the poles $c_1,c_2$. The contribution is $\nds \log
\left(c_2/c_1\right) $, which   can be seen from the first term  of $\dot{\varphi}_1$ in \eref{4.7}. Finishing 
all computations  we have 
\begin{align}
\Big( \braket{\dot{\varphi}_I|\varphi_J} \Big) = \nds \left(
\begin{array}{cc}
 \frac{\epsilon }{\delta } \log \left(c_1c_2\right) & \log
   \left(c_2/c_1\right) \\
 \log \left(c_2/c_1\right) & \frac{\delta }{\epsilon } \log \left(c_1c_2\right) + 4\log(c_1-c_2) \\
\end{array}
\right)  \, .
\end{align}
Then, we have
\begin{align}
\nds \Omega (c_i;\ep,\delta) = \nds \left(
\begin{array}{cc}
 \frac{\delta +2 \ep}{2} \log \left(c_1 c_2\right)& \frac{\epsilon}{2}   \log \left(\frac{c_2}{c_1}\right) \\
 \frac{\delta   (\delta +2 \epsilon )}{2 \epsilon } \log \left(\frac{c_2}{c_1}\right) 
 & \frac{\delta}{2}   \log \left(c_1 c_2\right)+2 \epsilon  \log \left(c_1-c_2\right) \\
\end{array}
\right)   \, . \label{eq:1varDEreg}
\end{align}
At this moment, one sees  that $\nds \Omega_{21} \neq 0$, which is not physical since  it means that the decomposition of  the differential of the subsector integral will rely on the higher sector. This is the consequence 
of involving a regulator, just as we have discussed in the last two sections. Take the regulator $\delta$ to be zero, we get the final result
\begin{align}
\nds \Omega (c_i;\ep,0) = \ep~  \nds \left(
\begin{array}{cc}
 \log \left(c_1 c_2\right) & \frac{1}{2} \log \left(c_2/c_1\right) \\
 0 & 2  \log
   \left(c_1-c_2\right) \\
\end{array}
\right)  \, . \label{eq:1varDEreg0}
\end{align}.

\subsection{With relative cohomology}

In this case, we have
\begin{align}
&u= (z-c_1)^\ep(z-c_2)^\ep
\, , \ \  D_1=z   \nn\\
&\omega=\nd \log u = \left( \frac{\ep}{z-c_1}+ \frac{\ep}{z-c_2} \right) \nd z \, .
\end{align}
The dual basis in relative cohomology is now $\Delta_{\hat{1}}\varphi_1, \Delta_{\hat{2}}\varphi_2$.  
Again we need to consider contributions from the following locations 
\begin{align}
\{\bm{p}\}=\{0,c_1,c_2,\infty\} \, .
\end{align}
Computations of $\braket{\varphi_I|\Delta_{\hat{1}}\varphi_1}$ will be changed due to cutting\footnote{For this example, $z$ corresponds to the propagator, so there is no cut available for $\varphi_2$, so we have  $\braket{\star|\Delta_{\hat{2}}\varphi_2}=\braket{\star|\varphi_2} $.} 
\begin{align}
&\braket{\varphi_1|\Delta_{\hat{1}}\varphi_1}=\braket{\varphi_{1;\hat{1}}|\varphi_{1;\hat{1}}} = \braket{1|1}=1  \nn\\
&\braket{\varphi_2|\Delta_{\hat{1}}\varphi_1}=\braket{\varphi_{2;\hat{1}}|\varphi_{1;\hat{1}}} = \braket{0|1}=0 \nn\\
& \braket{\varphi_1|\Delta_{\hat{2}}\varphi_2}=\braket{\varphi_1|\varphi_2}=0\nn \\
& \braket{\varphi_2|\Delta_{\hat{2}}\varphi_2}=\braket{\varphi_2|\varphi_2}={2\over \epsilon}~~~\label{4.16}
\end{align}
For $\braket{\dot{\varphi}_I|\Delta_{\hat{J}}\varphi_J}$, $\dot{\varphi}_1$ is same as the one given in \eref{4.7} 
while
\bea \dot{\varphi}_2 &= & \nD \log u \wedge \nD \log \left(z-c_1\right) - \nD \log u \wedge \nD \log \left(z-c_2\right) \nn\\
& = &2 \ep ~  \nD \log \left(z-c_2\right) \wedge \nD \log \left(z-c_1\right) \eea
thus we have
\begin{align}
&\braket{\dot{\varphi}_1|\Delta_{\hat{1}}\varphi_1}=\braket{\dot{\varphi}_{1;\hat{1}}|\varphi_{1;\hat{1}}} = \braket{\nds\log u|_{z=0}|1}= \ep \nds \log (c_1 c_2) \nn\\
&\braket{\dot{\varphi}_2|\Delta_{\hat{1}}\varphi_1}=\braket{\dot{\varphi}_{2;\hat{1}}|\varphi_{1;\hat{1}}} = \braket{0|1}= 0 \nn \\
& \braket{\dot{\varphi}_2|\Delta_{\hat{2}}\varphi_2}=\braket{\dot{\varphi}_2|\varphi_2} = 4\ep \nds \log(c_1-c_2)\nn \\
&\braket{\dot{\varphi}_1|\Delta_{\hat{2}}\varphi_2}=\braket{\dot{\varphi}_1|\varphi_2}=\ep \nds \log \left(c_2/c_1\right)
\end{align}
Putting all together we have  
\begin{align}
&\eta=\braket{ \varphi_{I;\hat{J}}|\varphi_{J;\hat{J}}} =\left(
\begin{array}{cc}
 1 & 0 \\
 0 & 2/\epsilon  \\
\end{array}
\right)  \, , \quad  \eta^{-1}=\left(
\begin{array}{cc}
 1  & 0 \\
 0 & \epsilon/2 \\
\end{array}
\right)  \, , \nn\\
&\Big( \braket{\dot{\varphi}_{I;\hat{J}}|\varphi_{J;\hat{J}}} \Big) = \nds \left(
\begin{array}{cc}
\ep \log \left(c_1 c_2\right) 
& \ep \log \left(c_2/c_1\right) \\
0 
& 4 \log
   \left(c_1-c_2\right) \\
\end{array}
\right)  \, \nn\\
&\nds \Omega =\Big( \braket{\dot{\varphi}_{I;\hat{J}}|\varphi_{J;\hat{J}}} \Big).\eta^{-1}= \ep~  \nds \left(
\begin{array}{cc}
 \log \left(c_1 c_2\right) & \frac{1}{2} \log \left(c_2/c_1\right) \\
 0 & 2  \log
   \left(c_1-c_2\right) \\
\end{array}
\right)  \, .
\end{align}
We get the same result as \eref{eq:1varDEreg0}, without extra terms in the intermediate steps \eref{eq:1varDEreg}. Cut also makes the computation easier.


\section{2-loop example: kite topology}\label{sec:eg2}

Now we consider the kite topology defined by
\begin{align}
    z_1&=l_1^2-m^2 \,,\; z_2=(l_2-p)^2-m^2 \,,\; z_3=(l_1-l_2)^2 \,,\nn
    \\
    z_4&=l_2^2 \,,\quad z_5=(l_1-p)^2 \,,\quad p^2=s \,.
\end{align}
with cut $z_1, z_2, z_3$.  We are going to compute it using the relative cohomology. For comparison, we also 
present the computation of $\eta$ using the regulator method. 

\subsection{Computation of  $\eta$ with regulator} 

In the Baikov representation, u of this integral family with regulator is
\begin{align}
& u(z_4,z_5) = z_4^{\delta_1} z_5^{\delta_2} \left[ \mG(z_4,z_5) \right]^{-\epsilon}  \nn\\
&\mG(z_4,z_5) \equiv 4\nG(l_1,l_2,p)|_{z_{1,2,3}=0} = - 2m^6 + m^4 (s + z_4 + z_5) \nn\\
& \quad \quad \quad \quad+ m^2 (2 z_4 z_5 - s z_4 - s z_5) + z_4 z_5 (s-z_4-z_5) \,,
\end{align}
where $\nG(l_1,l_2,p)$ is Gram determinant  defined by
\begin{align}
\nG(q_1,q_2,\cdots,q_n) \equiv |q_i\spdot q_j|
\end{align}
If ignoring the exchange symmetry $z_4 \leftrightarrow z_5$, there are four master integrals $f_i$ in this integral family and their integrands can be constructed as $\nd \log$-forms
\begin{align}
& f_i = \int u \varphi_i \nn\\
	&\varphi_{1}=\nd \log(z_4)\wedge\nd \log(z_5) =\frac{\nd z_4 \nd z_5}{z_4 z_5} \,, \nn\\
	&\varphi_{2}=\nd \log(\tau[z_4,m^2;r_{1;\pm}])\wedge\nd \log\left(\frac{z_5-r_{5+}}{z_5-r_{5-}}\right) =\frac{\sqrt{s(s-4m^2)}}{\mG}\nd z_4 \nd z_5 , \nn\\
	&\varphi_{3}=-\nd \log(\tau[z_4,\infty;r_{1;\pm}])\wedge\nd \log\left(\frac{z_5-r_{5+}}{z_5-r_{5-}}\right) =\frac{ z_4-m^2 }{\mG}\nd z_4 \nd z_5, \nn\\
	&\varphi_{4}=-\nd \log(\tau[z_5,\infty;r_{1;\pm}])\wedge\nd \log\left(\frac{z_4-r_{4+}}{z_4-r_{4-}}\right)=\frac{ z_5-m^2 }{\mG}\nd z_4 \nd z_5,  \label{eq:dlog}
\end{align}
where 
\begin{align}
\mG_1(z_5) &\equiv -4\nG(l_1,p)|_{z_{1,2,3}=0} = (z_5-s)^2+m^4-2m^2(z_5+s)
\end{align}
and  various roots of quadratic polynomials are defined by
\begin{align}  
&r_{\pm}[a x^2 + b x +c;x] \equiv \frac{-b\pm\sqrt{b^2-4ac}}{2a} \, , \nn\\
  &r_{1;\pm}\equiv  r_{\pm}[\mG_1(z_5);z_5]
  \,, \quad r_{4\pm}(z_5)\equiv r_{\pm}[\mG;z_4]\,, \quad r_{5\pm}(z_4)\equiv r_{\pm}[\mG;z_5]\,,  \,.\label{eq:polevalue}
\end{align}
One can see that with the above definition 
\bea r_{5+}(\infty)=\infty \,, \quad r_{5-}(\infty)=0 \,, \quad r_{5\pm}(m^2)=m^2\eea
Notice that if taking exchange symmetry $z_4 \leftrightarrow z_5$ into consideration, there are only three master integrals remaining and $\int u\varphi_3=\int u \varphi_4$.

For the $\nd \log$ basis we have chosen, the location  $(z_4,z_5)$ of poles are 
\begin{equation}
    \bm{p} \in \left\{ (0,0), (m^2,m^2), (\infty,0), (0,\infty), (\infty,\infty) \right\} ~~~\label{5.8}
\end{equation}
To describe the behavior around $z_4\to \infty$ and $z_5\to \infty$, we use variables $t_4=1/z_4$ and  $t_5=1/z_5$.
With the changing of variables, $u$ around poles $(\infty,0), (0,\infty), (\infty,\infty)$ becomes
\bea
u_{\infty 0}=t_4^{2\epsilon-\delta_1} z_5^{\delta_2} \, \mG_{\infty 0}^{-\epsilon},~~~
u_{0 \infty}= z_4^{\delta_2}t_5^{2\epsilon-\delta_2} \, \mG_{ 0\infty}^{-\epsilon},~~~~
u_{\infty \infty}=t_4^{2\epsilon-\delta_1} t_5^{2\ep-\delta_2} \, \mG_{\infty \infty}^{-\epsilon} \eea
where 
\begin{align}
&\mG_{\infty 0}(t_4,z_5) \equiv t_4^2\  \mG\left(1/t_4,z_5\right)=-2 m^6 t_4^2 +m^4 s t_4^2+m^4 t_4 
+m^4 t_4^2 z_5-m^2 s t_4 \nn\\ 
& \quad \quad \quad \quad \quad  -m^2 s t_4^2 z_5+2 m^2 t_4 z_5+s t_4 z_5 -t_4 z_5^2-z_5 \,,\nn\\
& \mG_{0\infty }(z_4,t_5) \equiv t_5^2\  \mG\left(z_4,1/t_5\right) =\mG_{\infty 0}(t_4 \to t_5, z_5 \to z_4) \,.  \nn\\
&\mG_{\infty \infty }(t_4,t_5) \equiv t_4^2t_5^2\  \mG\left(1/t_4,1/t_5\right) = (-2 m^6 +m^4 s )t_4^2 t_5^2 
+(m^4 -m^2 s) t_4 t_5^2 \nn\\ 
& \quad \quad  \quad \quad \quad + (m^4 -m^2 s) t_4^2 t_5+2 m^2 t_4 t_5 
+s t_4 t_5-t_4-t_5 \,,
\label{eq:Ginf}
\end{align}
Let us analysis the pole $t_4=0,z_5=0$ for $u_{\infty 0}$. From the explicit expression of $\mG_{\infty 0}(t_4,z_5) $, we can see the leading terms are $m^4 t_4-m^2 s t_4-z_5$. Thus three hypersurfaces $z_5=0$, $t_4=0$ and $m^4 t_4-m^2 s t_4-z_5$ meet at $(z_4,z_5)=(\infty,0)$, so $(\infty,0)$ is a degenerate pole. Using the 
relation $\mG_{0\infty }(z_4,t_5)=\mG_{\infty 0}(t_4 \to t_5, z_5 \to z_4)$, we see that $(z_4,z_5)=(0,\infty)$ is a degenerate pole. Finally, the leading term  of  $\mG_{\infty \infty }(t_4,t_5) $ is $-t_4-t_5$,
so $(z_4,z_5)=(\infty,\infty)$ is a degenerate pole. 
 In \cite{Chen:2023kgw}, we have studied polynomial $\mG$ around each pole using  the Newton polytopes of $\mG$,
 where each monomial corresponds to a vertex of the polytope. Here we will not discuss this point further.

With the above analysis, for poles given in \eref{5.8}, there are a total $12$ contours used for the computation. For the last three degenerated poles, each 
pole has used $3$ contours. The residue of pole $(m^2,m^2)$ is the composite residue \cite{Arkani-Hamed:2009ljj,tsikh1992multidimensional,auizenberg1983integral}, for which we use two contours to compute it. Now we list the contour and the corresponding factorization transformations for the computation of residue
when computing the intersection number. 
For $\bm{p}=(0,0)$, it does not need the factorization. However, for the sake of formal uniformity, we
 denote identity transformation as  $\nT^{(1)}$ 
\begin{align}
\nT^{(1)} &: (\{z_4\},\{z_5\})  \, , \ \  z_4\to x^{(1)}_1 \, , \ z_5\to x^{(1)}_2 \,,  \label{eq:T1reg}
\end{align}
For $\bm{p}=(m^2,m^2)$, involved contours and factorizations could be chosen as
\begin{align}
\nT^{(2)} &: (\{z_4-m^2,z_5-r_{5+}\} , \{z_5-r_{5-}\}),\nn\\
&~ z_4 \to x_1^{(2)} x_2^{(2)} +r_{+}[\mG(x_1^{(2)},x_2^{(2)});x_1^{(2)}] \,, \quad z_5 \to x_2^{(2)} \,,\nn\\
\nT^{(3)} &: (\{z_4-m^2,z_5-r_{5-}\} , \{z_5-r_{5+}\}),\nn\\
&~ z_5 \to x_1^{(3)} x_2^{(3)} +r_{+}[\mG(x_1^{(3)},x_2^{(3)});x_2^{(3)}] \,, \quad z_4 \to x_2^{(3)} \,.\label{eq:T23reg}
\end{align}
For $\bm{p}=(\infty,0)$, they are 
 \begin{align}
\nT^{(4)} &: (\{t_4\} , \{z_5,z_5-r_-[\mG_{\infty0};z_5]\}),,\nn\\
&~t_4 \to x_1^{(4)} x_2^{(4)} \,, \quad z_5 \to x_2^{(4)} \,, \nn
\\
\nT^{(5)} &: (\{t_4,z_5-r_-[\mG_{\infty0};z_5]\},\{z_5\}),\nn\\
&~ t_4 \to x_1^{(5)} \,, \quad z_5 \to x_1^{(5)} x_2^{(5)} \,, \nn
\\
\nT^{(6)} &: (\{z_5-r_-[\mG_{\infty0};z_5] \}, \{t_4,z_5\}), \nn\\
&~t_4 \to x_1^{(6)} x_2^{(6)} + r_{+}[\mG_{\infty0}(x_1^{(6)},x_2^{(6)});x_1^{(6)}] \,, \quad z_5\to  x_2^{(6)} \,.
\end{align}
For $\bm{p}=(0,\infty)$, contours are
 \begin{align}
\nT^{(7)} &: (\{t_5\} , \{z_4,z_4-r_-[\mG_{0\infty};z_4]\}),\nn\\
\nT^{(8)} &: (\{t_5,z_4-r_-[\mG_{0\infty};z_4]\}\{z_4\} ),\nn\\
\nT^{(9)} &: (\{z_4-r_-[\mG_{0\infty};z_4]\} , \{z_4,t_5\}).
\end{align}
and their factorization could be obtained from $\nT^{(4,5,6)}$ via  $z_4 \leftrightarrow z_5$ symmetry.
For $\bm{p}=(\infty,\infty)$, contours and factorizations are
 \begin{align}
\nT^{(10)} &: (\{t_4\} , \{t_5,t_4-r_+[\mG_{\infty\infty};t_4]\}),\nn\\
&~t_4 \to x_1^{(10)} x_2^{(10)} \,, \quad t_5\to x_2^{(10)} \,,
\\
\nT^{(11)} &: ( \{t_4,t_4-r_+[\mG_{\infty\infty};t_4]\},\{t_5\} ),\nn\\
&~t_4 \to  x_1^{(11)} \,, \quad t_5\to x_1^{(11)} x_2^{(11)} \,,
\\
\nT^{(12)} &: (\{t_4-r_+[\mG_{\infty\infty};t_4]\} , \{t_4,t_5\}) \nn\\
&~ t_4 \to x_1^{(12)} x_2^{(12)} + r_{+}[\mG_{\infty\infty}(x_1^{(12)},x_2^{(12)});x_1^{(12)}] \,, \quad t_5\to x_2^{(12)} \,.\label{eq:T101112reg}
\end{align}
From these factorizations, we can read out hypersurface powers as
\begin{align}
&\bm{\gamma}^{(1)}=\delta_1 \delta_2 \,, \quad \bm{\gamma}^{(2,3)}=(-2\ep)(-\ep) \,, \quad \bm{\gamma}^{(7,8,9)}= \bm{\gamma}^{(4,5,6)}\Big|_{\delta_1 \leftrightarrow \delta_2} \,,\nn\\
&\bm{\gamma}^{(4)}= \left(\ep-\delta _1+\delta _2\right)\left(2 \ep -\delta_1\right) \, , \quad 
 ~~~~~~\bm{\gamma}^{(10)}=\left(3\ep-\delta _1-\delta _2\right) \left(2 \ep -\delta_1\right),\nn\\
 &\bm{\gamma}^{(5)}= \left(\ep-\delta _1+\delta _2\right)\delta_2 \,, \quad ~~~~~~~~~~~~~~~\bm{\gamma}^{(11)}=\left(3\ep-\delta _1-\delta _2\right) \left(2 \ep -\delta_2\right),\nn\\
 &\bm{\gamma}^{(6)}=\left(\ep-\delta _1+\delta _2\right)(-\ep ) \,, \quad  
 ~~~~~~~~~~~~\bm{\gamma}^{(12)}=\left(3\ep-\delta _1-\delta _2\right)(-\ep) \,.~~~\label{5.18}
\end{align}
One can notice that the  hypersurface power $\beta_i(\beta_1+\beta_2+\beta_3)$ in  \eref{eq:examfactorize} appears here for these degenerate poles again, since they have the same degenerate structure (two dimension, three hypersurfaces).
$C^{(\bm{-1})}_I$ for each factorization are
\begin{align}
&\varphi_{1}:\{1,0,0,-1,1,0,-1,1,0,1,-1,0\}\,,\nn\\
&\varphi_{2}:\{0,-1,-1,0,0,0,0,0,0,0,0,0\}\,,\nn\\
&\varphi_{3}:\{0,0,0,1,0,-1,0,0,0,-1,0,1\}\,,\nn\\
&\varphi_{4}:\{0,0,0,0,0,0,1,0,-1,0,1,-1\}\,.  \label{eq:commonpole2l}
\end{align}
We see that given a basis $\varphi_{1}$ some contours give a nonzero contribution and some contours,  zero.
This phenomenon can be easily understood.  First, the basis should have the corresponding pole for the contour.
Secondly, the poles should be properly grouped. For example, $\varphi_3$ at pole $(\infty,0)$ has two factors  $t_4$ and $z_5-r_-[\mG_{\infty0};z_5]$ in the denominator, thus only the contour with the grouping 
$(\{t_4,\star\},\{z_5-r_-[\mG_{\infty0};z_5],\star\})$ ( the $\star$ could be no element or several elements)
can have a nonzero contribution. One can see that $\nT^{(4)}$ and $\nT^{(6)}$ satisfy these conditions, thus the related $C^{(\bm{-1})}_3$ is non-zero.
Similar understanding for other bases.

Collecting everything together by \eref{eq:LO} we have
\begin{align} 
&\eta=\left(
\begin{array}{cccc}
 \sum_{\alpha=1,4,5,7,8,10,11}\frac{1}{\bm{\gamma}^{(\alpha)}} & 0 & -\frac{1}{\bm{\gamma}^{(4)}}-\frac{1}{\bm{\gamma}^{(10)}} & -\frac{1}{\bm{\gamma}^{(7)}}-\frac{1}{\bm{\gamma}^{(11)}} \\
 0 & \frac{1}{\bm{\gamma}^{(2)}}+\frac{1}{\bm{\gamma}^{(3)}} & 0 & 0 \\
 -\frac{1}{\bm{\gamma}^{(4)}}-\frac{1}{\bm{\gamma}^{(10)}} & 0 &  \sum_{\alpha=4,6,10,12}\frac{1}{\bm{\gamma}^{(\alpha)}} & -\frac{1}{\bm{\gamma}^{(12)}} \\
 -\frac{1}{\bm{\gamma}^{(7)}}-\frac{1}{\bm{\gamma}^{(11)}} & 0 & -\frac{1}{\bm{\gamma}^{(12)}}  &  \sum_{\alpha=7,9,11,12}\frac{1}{\bm{\gamma}^{(\alpha)}} \\
\end{array}
\right) \nn\\
&=\left(
\begin{array}{cccc}
 \frac{\epsilon  \left(2 \delta _1 \epsilon +2 \delta _2 \epsilon -\delta _1^2-\delta _2^2+2
   \delta _1 \delta _2+3 \epsilon ^2\right)}{\delta _1 \delta _2 \left(-\delta _1-\delta _2+3
   \epsilon \right) \left(\delta _1-\delta _2+\epsilon \right) \left(-\delta _1+\delta
   _2+\epsilon \right)} & 0 & -\frac{2}{\left(-\delta _1-\delta _2+3 \epsilon \right)
   \left(-\delta _1+\delta _2+\epsilon \right)} & -\frac{2}{\left(-\delta _1-\delta _2+3
   \epsilon \right) \left(\delta _1-\delta _2+\epsilon \right)} \\
 0 & \frac{1}{\epsilon ^2} & 0 & 0 \\
 -\frac{2}{\left(-\delta _1-\delta _2+3 \epsilon \right) \left(-\delta _1+\delta _2+\epsilon
   \right)} & 0 & -\frac{2 \left(\epsilon -\delta _1\right)}{\epsilon  \left(-\delta _1-\delta
   _2+3 \epsilon \right) \left(-\delta _1+\delta _2+\epsilon \right)} & \frac{1}{\epsilon 
   \left(-\delta _1-\delta _2+3 \epsilon \right)} \\
 -\frac{2}{\left(-\delta _1-\delta _2+3 \epsilon \right) \left(\delta _1-\delta _2+\epsilon
   \right)} & 0 & \frac{1}{\epsilon  \left(-\delta _1-\delta _2+3 \epsilon \right)} & -\frac{2
   \left(\epsilon -\delta _2\right)}{\epsilon  \left(-\delta _1-\delta _2+3 \epsilon \right)
   \left(\delta _1-\delta _2+\epsilon \right)} \\
\end{array}
\right)~~~\label{5.20}
\end{align}
and
\begin{align}
\eta^{-1} = \begin{pmatrix}
	\frac{\delta _1 \delta _2 \left(-\delta _1-\delta _2+\epsilon \right)}{\epsilon } & 0 & -2 \delta _1 \delta _2 & -2 \delta _1 \delta _2 \\
	0 & \epsilon ^2 & 0 & 0 \\
	-2 \delta _1 \delta _2 & 0 & -2 \epsilon  \left(\delta _2+\epsilon \right) & -\epsilon  \left(\delta _1+\delta _2+\epsilon \right) \\
	-2 \delta _1 \delta _2 & 0 & -\epsilon  \left(\delta _1+\delta _2+\epsilon \right) & -2 \epsilon  \left(\delta _1+\epsilon \right) \\
	\end{pmatrix} . \label{eq:invetareg2}
\end{align}


\subsection{Computation of  $\eta$ with relative cohomology}
In this case, we have
\begin{align}
& u(z_4,z_5) = \left[ \mG(z_4,z_5) \right]^{-\epsilon} \,, \ \ D_1=z_4\, , \ D_2=z_5 \, ,  \nn\\
&u_{\infty 0}=t_4^{2\epsilon} \, \mG_{\infty 0}^{-\epsilon} \, , \quad u_{0 \infty}= t_5^{2\epsilon}\mG_{0\infty }^{-\epsilon} \, , \quad  u_{\infty \infty}=t_4^{2\epsilon} t_5^{2\ep} \, \mG_{\infty \infty}^{-\epsilon} 
\end{align}
where $\mG$'s are given in  \eref{eq:Ginf}.
Using the same $\nd \log$ integrand we have contructed in \eref{eq:dlog}, again we need to consider poles located at 
\begin{equation}
    \bm{p} \in \left\{ (0,0), (m^2,m^2), (\infty,0), (0,\infty), (\infty,\infty) \right\} 
\end{equation}
For the pole $(0,0)$ it can only contribute to the top sector. The contour $\nT^{(1)}$ in previous subsection should be replaced by the contour $\nT^{(13)}$, which transform nothing, i.e., $\nT^{(13)} : ()$. The evaluation 
of intersection numbers is just like the first line of  \eref{4.16}, i.e., constant to constant.

For poles $(\infty,0)$ and $(0,\infty)$, the discussion is a little bit tricky. When we consider $\braket{\varphi_{I;\hat{J}}|\varphi_{J;\hat{J}}},\  I,J=2,3,4$, since there is no $z_5$ in $u_{\infty 0 }$ and no $z_4$ in $u_{0 \infty }$, they are no longer degenerate poles  between subsectors $\mathcal{B}_{\hat{2}}=\mathcal{B}_{\hat{3}}=\mathcal{B}_{\hat{4}}$ for basis $\varphi_{2}, \varphi_{3},\varphi_{4}$ (given in \eref{eq:dlog}) containing no $z_4, z_5$. 
Thus when compute the intersection number, for pole $(\infty,0)$ we can use  the contour 
and the factorization 
 \begin{align}
\nT^{(14)} &: (\{t_4\} , \{z_5-r_-[\mG_{\infty0};z_5]\}),,\nn\\
&~t_4 \to x_1^{(4)} \,, \quad z_5\to  x_2^{(14)} +  r_{+}[\mG_{\infty0}(x_1^{(14)},x_2^{(14)});x_2^{(14)}] \,.
\end{align}
and  for pole $(0,\infty)$, the contour and factorization   
\begin{align}
\nT^{(15)} &: (\{t_5\} , \{z_4-r_-[\mG_{\infty0};z_4]\}),,\nn\\
&~t_5 \to x_1^{(15)} \,, \quad z_4\to  x_2^{(15)} +  r_{+}[\mG_{0\infty}(x_1^{(15)},x_2^{(15)});x_2^{(15)}] \,.
\end{align}
One can check that the contributions of  $\nT^{(14)}$ and $\nT^{(15)}$ can be obtained via sum the contribution of $\nT^{(4,6)}$
and $\nT^{(7,9)}$ respectively, for example, $1/\bm{\gamma}^{(14)} = 1/\bm{\gamma}^{(4)} +1/\bm{\gamma}^{(6)}$, which is nothing, but the \eref{eq:regionrule} in previous section.

However, when we compute $\braket{\varphi_{1;\hat{J}}|\varphi_{J;\hat{J}}},\  J=2,3$, since $\varphi_{1}$ contains
the denominator $z_4z_5$, poles $(\infty,0)$ and $(0,\infty)$ are still degenerate and we should use the 
contours $\nT^{(4,5,6)}$
and $\nT^{(7,8,9)}$. From \eref{eq:commonpole2l} one can see that the fifth and the eighth components of vector $C^{(\bm{-1})}_I$ of $\varphi_{2,3,4}$ are zero, so the contributions from $\nT^{(5,8)}$ are zero. 

The above two situations can be combined into the statement that we should use the contours $\nT^{(4,6,7,9)}$ for the 
computation of all intersection numbers. 
Then, all the hypersurface powers are
\begin{align}
&\bm{\gamma}^{(13)}=1 \,, \quad \bm{\gamma}^{(2,3)}=(-2\ep)(-\ep) \,, \quad \bm{\gamma}^{(4)}= \bm{\gamma}^{(7)}= 2\ep(\ep) \,,\quad \bm{\gamma}^{(6)}= \bm{\gamma}^{(9)}= -\ep(\ep) \,,\nn\\
&\bm{\gamma}^{(10)}=3\ep \left(2 \ep\right) \, , \quad \bm{\gamma}^{(11)}=3\ep \left(2 \ep\right) \, ,
\quad \bm{\gamma}^{(12)}=3\ep(-\ep) \,.
\end{align}
The $C^{(\bm{-1})}_I$ for $\nT^{(13,2,3,4,6,7,9,10,11,12)}$ are
\begin{align}
&\varphi_{1}:\{1,0,0,-1,0,-1,0,1,-1,0\}\,,\nn\\
&\varphi_{2}:\{0,-1,-1,0,0,0,0,0,0,0\}\,,\nn\\
&\varphi_{3}:\{0,0,0,1,-1,0,0,-1,0,1\}\,,\nn\\
&\varphi_{4}:\{0,0,0,0,0,1,-1,0,1,-1\}\,.  \label{eq:commonpoleRC}
\end{align}
Using \eref{eq:LO} one could easily compute $\eta$ and $\eta^{-1}$ as
\begin{align} 
&\eta=\left(
\begin{array}{cccc}
 1 & 0 & -\frac{1}{\bm{\gamma}^{(4)}}-\frac{1}{\bm{\gamma}^{(10)}} & -\frac{1}{\bm{\gamma}^{(7)}}-\frac{1}{\bm{\gamma}^{(11)}} \\
 0 & \frac{1}{\bm{\gamma}^{(2)}}+\frac{1}{\bm{\gamma}^{(3)}} & 0 & 0 \\
0 & 0 &  \sum_{\alpha=4,6,10,12}\frac{1}{\bm{\gamma}^{(\alpha)}} & -\frac{1}{\bm{\gamma}^{(12)}} \\
0 & 0 & -\frac{1}{\bm{\gamma}^{(12)}}  &  \sum_{\alpha=7,9,11,12}\frac{1}{\bm{\gamma}^{(\alpha)}} \\
\end{array}
\right)=\left(
\begin{array}{cccc}
 1 & 0 & -\frac{2}{3 \ep^2} & -\frac{2}{3 \ep^2} \\
 0 & \frac{1}{\epsilon ^2} & 0 & 0 \\
 0 & 0 & -\frac{2}{3 \ep^2} & \frac{1}{3 \ep^2} \\
0 & 0 & \frac{1}{3 \ep^2} & -\frac{2}{3 \ep^2} \\
\end{array}
\right)  \nn\\
&\eta^{-1}= \left(
\begin{array}{cccc}
 1 & 0 & -2 & -2 \\
 0 & \epsilon ^2 & 0 & 0 \\
 0 & 0 & -2 \epsilon ^2 & -\epsilon ^2 \\
 0 & 0 & -\epsilon ^2 & -2 \epsilon ^2 \\
\end{array}
\right)~~~~\label{5.28}
\end{align}
Compared to the  computation \eref{5.20} with regulator, one could see that we  have simpler expressions in intermediate steps.


\subsection{$\nd \Omega_{1J}$ with relative cohomology}

$\varphi_{2,3,4}$ belong to subsector and their loop-by-loop Baikov representation is an univariate problem with $u=(z-m^2)^{-2\ep}z^{\ep}(z^2+m^4+s^2-2zm^2-2sm^2-2zs)^{-1/2-\ep}$. Since such cases have been systematically discussed and are easily computed using techniques  for univariate problems, we do not compute them using 2-variables $u(z_4,z_5)$ here. Hence in this subsection, we compute $\nd \Omega_{1J}$ only. 

For $\nd \Omega_{11}$, from the non zero terms in $\left(\eta^{-1}\right)_{J1}$ \eref{5.28},  only $\braket{\dot{\varphi}_{1;\hat{1}}|\varphi_{1;\hat{1}}}\left(\eta^{-1}\right)_{11}$ could contribute. Thus we only need to compute
\begin{align} 
\nd \Omega_{11} = \braket{\dot{\varphi}_{1;\hat{1}}|\varphi_{1;\hat{1}}} = \nds \log u(0,0) =-\ep  (2\nds \log m^2 + \nds \log (s-2m^2) )  \,. \label{eq:d11}
\end{align}
where \eref{tot-D} has been used for the maximal cut $z_4=z_5=0$, i.e.,
\begin{align}
\nds u(0,0) = (\nds \Omega)~  u(0,0) \ \to \  \nds \Omega = \nds \log u(0,0)
\end{align}
as a trivial 0-variable problem. If we compute with regulator, it will be more complicated,  since by \eref{eq:invetareg2}, $\left(\eta^{-1}\right)_{11}$, $\left(\eta^{-1}\right)_{31}$ and $\left(\eta^{-1}\right)_{41}$ are all non-zero and the corresponding computations could not apply maximal cut as did for $\mathcal{B}_{\hat{1}}$. 

Now we move to $\nd \Omega_{1J}$, $J=2,3,4$. To get $\nd \Omega_{12}$, we only need compute $\braket{\dot{\varphi}_{1;\hat{2}}|\varphi_{2;\hat{2}}}$. However, $\varphi_1$ and $\varphi_2$  do not share 2-SP or 1-SP. We have
\begin{align} 
\nd \Omega_{12} = \ep^2 \braket{\dot{\varphi}_{1;\hat{2}}|\varphi_{2;\hat{2}}} = 0
\end{align}
To get $\nd \Omega_{13}$, we need to compute $\braket{\dot{\varphi}_{1;\hat{J}}|\varphi_{J;\hat{J}}}$, $J=1,3,4$. $J=1$ has been computed in \eref{eq:d11}. Due to the $z_4\leftrightarrow z_5$ symmetry, $\braket{\dot{\varphi}_{1;\hat{3}}|\varphi_{3;\hat{3}}}= \braket{\dot{\varphi}_{1;\hat{4}}|\varphi_{3;\hat{4}}}$.
For $\braket{\dot{\varphi}_{1;\hat{3}}|\varphi_{3;\hat{3}}}$, $\nT^{(4,10)}$ lead to 2-SP, whose contributions vanish due to $\nds \log C = 0$. $\nT^{(11,12)}$ lead to shared 1-SP, whose contributions also vanish due to $\nds \log C = 0$. Only $\nT^{(5,6)}$ lead to the same shared 1-SP $(\{t_4,\star\},\{\star\})$ which give the only non-zero contribution
\begin{align}
\braket{\dot{\varphi}_{1;\hat{3}}|\varphi_{3;\hat{3}}}=\frac{1}{\ep} \left( \nds \log(m^2) + \nds \log(s-m^2) \right) \, . \label{eq:NPTCLetter}
\end{align}
Then we have
\begin{align} 
\nd \Omega_{13}&=\left(\eta^{-1}\right)_{13} \braket{\dot{\varphi}_{1;\hat{1}}|\varphi_{1;\hat{1}}}  +  \left[\left(\eta^{-1}\right)_{33}+\left(\eta^{-1}\right)_{43}\right)]\braket{\dot{\varphi}_{1;\hat{3}}|\varphi_{3;\hat{3}}}  \nn\\
&= (-2)( -\ep)\nds (2  \log (m^2) +  \log (s-2m^2) )+ (-2 \ep^2 - \ep^2)\frac{1}{\ep} \nds \left( \log(m^2) + \log(s-m^2) \right)\nn\\
&=\ep ~ \nds (\log (m^2)+ 2\log (s-2m^2) - 3 \log(s-m^2)         )
\end{align}
Due to the  $z_4\leftrightarrow z_5$ symmetry, we have
\begin{align} 
\nd \Omega_{13} =\nd \Omega_{14}\, . 
\end{align}
Now we have all $\nds \Omega_{1J}$.


\section{Summary and Outlook}\label{final}

In this paper, we focus on the computation of the CDE matrix with $\nd \log$-form basis ${\varphi}$. By writing the $\dot{\varphi}$ in $\nD \log$-form \eref{tot-D}, all computations of intersection numbers can be reduced to
the  LO contribution form given in \eref{eq:LO}. Thus we can show that the CDE matrix is the $\nds \log$-form
and if the powers of $u$ are proportional to $\ep$, then the differential equation is canonical. We have also shown
that  relative cohomology can simplify the computation of intersection numbers by giving a simple computation rule. Our selection rule, i.e., n-SP and (n-1)-SP contributions, has provided better insights into CDE. We  provide also careful treatment on the factorization of multivariate poles. 
To demonstrate the utility of the above analysis, we have presented detailed computations using two examples.  
Our results will help the understanding and application of intersection numbers.

To apply results in this paper to more complex examples, it necessitates an automatic and efficient algorithm for identifying all regions of poles and their factorization, which remains an open question to the best of our knowledge. However, our application of the analysis  does not need to be so rigid. 
For instance, if we care about the symbol only, the factorization transformations corresponding to the contours of all multivariable poles do not need to be complete, i.e., like the case \eqref{eq:overfactorized1} without 
getting the full region. The even worse case is that we do not have dlog integrand basis, but the analysis in this paper can still provide us  information about alphabet $\text{W}^{(i)}(\bm{s})$. For example, one could apply one of the factorization transformation
\begin{equation}
	u(\nT^{(\alpha)}[\bm{z}]) = \bar{u}_\alpha(\bm{x}) \prod_i \left[ x_i^{(\alpha)} - \rho_i^{(\alpha)} \right]^{\gamma_i^{(\alpha)}} ,
\end{equation}
and take $n-1$ variables to  values of poles to arrive at a univariate form (without loss of generality, we choose $x_1^{(\alpha)}$)
\begin{equation}
	u_1^{(\alpha)} \equiv \bar{u}_\alpha\left(x_1^{(\alpha)},\rho_2^{(\alpha)},\rho_3^{(\alpha)},\cdots  \right)  \left[ x_1^{(\alpha)} - \rho_1^{(\alpha)} \right]^{\gamma_1^{(\alpha)}} .
\end{equation}
As detail discussed in \cite{Chen:2023kgw}, the symbol letters of the univariate problem (if not elliptic) could be read out immediately from u. These letters are also the letters that could appear in the full multivariate problem. Thus, figure out $u_i^{(\alpha)}$ for all $\alpha$ and $i$ could 
provide letters even without the construction of $\nd \log$ integrand. These less stringent problems will be explored in the future.

\section*{Acknowledgements}
We thank Andrzej Pokraka for valuable discussions. This work is supported by Chinese NSF funding under Grant No.11935013, No.11947301, No.12047502 (Peng Huanwu Center), No.12247120, No.12247103, NSAF grant No.U2230402, and China Postdoctoral Science Foundation No.2022M720386.


\bibliographystyle{JHEP}
\bibliography{intnum_symbol}

\providecommand{\href}[2]{#2}\begingroup\raggedright\begin{thebibliography}{10}

\bibitem{Chetyrkin:1981qh}
K.G.~Chetyrkin and F.V.~Tkachov, \emph{{Integration by Parts: The Algorithm to
  Calculate beta Functions in 4 Loops}},
  \href{https://doi.org/10.1016/0550-3213(81)90199-1}{\emph{Nucl. Phys. B}
  {\bfseries 192} (1981) 159}.

\bibitem{Kotikov:1990kg}
A.V.~Kotikov, \emph{{Differential equations method: New technique for massive
  Feynman diagrams calculation}},
  \href{https://doi.org/10.1016/0370-2693(91)90413-K}{\emph{Phys. Lett. B}
  {\bfseries 254} (1991) 158}.

\bibitem{Kotikov:1991pm}
A.V.~Kotikov, \emph{{Differential equation method: The Calculation of N point
  Feynman diagrams}},
  \href{https://doi.org/10.1016/0370-2693(91)90536-Y}{\emph{Phys. Lett. B}
  {\bfseries 267} (1991) 123}.

\bibitem{Gehrmann:1999as}
T.~Gehrmann and E.~Remiddi, \emph{{Differential equations for two loop four
  point functions}},
  \href{https://doi.org/10.1016/S0550-3213(00)00223-6}{\emph{Nucl. Phys. B}
  {\bfseries 580} (2000) 485}
  [\href{https://arxiv.org/abs/hep-ph/9912329}{{\ttfamily hep-ph/9912329}}].

\bibitem{Bern:1993kr}
Z.~Bern, L.J.~Dixon and D.A.~Kosower, \emph{{Dimensionally regulated pentagon
  integrals}}, \href{https://doi.org/10.1016/0550-3213(94)90398-0}{\emph{Nucl.
  Phys. B} {\bfseries 412} (1994) 751}
  [\href{https://arxiv.org/abs/hep-ph/9306240}{{\ttfamily hep-ph/9306240}}].

\bibitem{Henn:2013pwa}
J.M.~Henn, \emph{{Multiloop integrals in dimensional regularization made
  simple}}, \href{https://doi.org/10.1103/PhysRevLett.110.251601}{\emph{Phys.
  Rev. Lett.} {\bfseries 110} (2013) 251601}
  [\href{https://arxiv.org/abs/1304.1806}{{\ttfamily 1304.1806}}].

\bibitem{Chen:1977oja}
K.-T.~Chen, \emph{{Iterated path integrals}},
  \href{https://doi.org/10.1090/S0002-9904-1977-14320-6}{\emph{Bull. Am. Math.
  Soc.} {\bfseries 83} (1977) 831}.

\bibitem{Goncharov:1998kja}
A.B.~Goncharov, \emph{{Multiple polylogarithms, cyclotomy and modular
  complexes}}, \href{https://doi.org/10.4310/MRL.1998.v5.n4.a7}{\emph{Math.
  Res. Lett.} {\bfseries 5} (1998) 497}
  [\href{https://arxiv.org/abs/1105.2076}{{\ttfamily 1105.2076}}].

\bibitem{Caron-Huot:2011dec}
S.~Caron-Huot and S.~He, \emph{{Jumpstarting the All-Loop S-Matrix of Planar
  N=4 Super Yang-Mills}},
  \href{https://doi.org/10.1007/JHEP07(2012)174}{\emph{JHEP} {\bfseries 07}
  (2012) 174} [\href{https://arxiv.org/abs/1112.1060}{{\ttfamily 1112.1060}}].

\bibitem{Golden:2013xva}
J.~Golden, A.B.~Goncharov, M.~Spradlin, C.~Vergu and A.~Volovich,
  \emph{{Motivic Amplitudes and Cluster Coordinates}},
  \href{https://doi.org/10.1007/JHEP01(2014)091}{\emph{JHEP} {\bfseries 01}
  (2014) 091} [\href{https://arxiv.org/abs/1305.1617}{{\ttfamily 1305.1617}}].

\bibitem{Panzer:2014caa}
E.~Panzer, \emph{{Algorithms for the symbolic integration of hyperlogarithms
  with applications to Feynman integrals}},
  \href{https://doi.org/10.1016/j.cpc.2014.10.019}{\emph{Comput. Phys. Commun.}
  {\bfseries 188} (2015) 148}
  [\href{https://arxiv.org/abs/1403.3385}{{\ttfamily 1403.3385}}].

\bibitem{Dennen:2015bet}
T.~Dennen, M.~Spradlin and A.~Volovich, \emph{{Landau Singularities and
  Symbology: One- and Two-loop MHV Amplitudes in SYM Theory}},
  \href{https://doi.org/10.1007/JHEP03(2016)069}{\emph{JHEP} {\bfseries 03}
  (2016) 069} [\href{https://arxiv.org/abs/1512.07909}{{\ttfamily
  1512.07909}}].

\bibitem{Caron-Huot:2016owq}
S.~Caron-Huot, L.J.~Dixon, A.~McLeod and M.~von Hippel, \emph{{Bootstrapping a
  Five-Loop Amplitude Using Steinmann Relations}},
  \href{https://doi.org/10.1103/PhysRevLett.117.241601}{\emph{Phys. Rev. Lett.}
  {\bfseries 117} (2016) 241601}
  [\href{https://arxiv.org/abs/1609.00669}{{\ttfamily 1609.00669}}].

\bibitem{Mago:2020kmp}
J.~Mago, A.~Schreiber, M.~Spradlin and A.~Volovich, \emph{{Symbol alphabets
  from plabic graphs}},
  \href{https://doi.org/10.1007/JHEP10(2020)128}{\emph{JHEP} {\bfseries 10}
  (2020) 128} [\href{https://arxiv.org/abs/2007.00646}{{\ttfamily
  2007.00646}}].

\bibitem{Abreu:2021vhb}
S.~Abreu, R.~Britto, C.~Duhr, E.~Gardi and J.~Matthew, \emph{{The diagrammatic
  coaction beyond one loop}},
  \href{https://doi.org/10.1007/JHEP10(2021)131}{\emph{JHEP} {\bfseries 10}
  (2021) 131} [\href{https://arxiv.org/abs/2106.01280}{{\ttfamily
  2106.01280}}].

\bibitem{Gong:2022erh}
J.~Gong and E.Y.~Yuan, \emph{{Towards analytic structure of Feynman parameter
  integrals with rational curves}},
  \href{https://doi.org/10.1007/JHEP10(2022)145}{\emph{JHEP} {\bfseries 10}
  (2022) 145} [\href{https://arxiv.org/abs/2206.06507}{{\ttfamily
  2206.06507}}].

\bibitem{Yang:2022gko}
Q.~Yang, \emph{{Schubert problems, positivity and symbol letters}},
  \href{https://doi.org/10.1007/JHEP08(2022)168}{\emph{JHEP} {\bfseries 08}
  (2022) 168} [\href{https://arxiv.org/abs/2203.16112}{{\ttfamily
  2203.16112}}].

\bibitem{He:2021non}
S.~He, Z.~Li and Q.~Yang, \emph{{Truncated cluster algebras and Feynman
  integrals with algebraic letters}},
  \href{https://doi.org/10.1007/JHEP12(2021)110}{\emph{JHEP} {\bfseries 12}
  (2021) 110} [\href{https://arxiv.org/abs/2106.09314}{{\ttfamily
  2106.09314}}].

\bibitem{He:2021eec}
S.~He, Z.~Li and Q.~Yang, \emph{{Kinematics, cluster algebras and Feynman
  integrals}},  \href{https://arxiv.org/abs/2112.11842}{{\ttfamily
  2112.11842}}.

\bibitem{He:2022tph}
S.~He, J.~Liu, Y.~Tang and Q.~Yang, \emph{{The symbology of Feynman integrals
  from twistor geometries}},
  \href{https://arxiv.org/abs/2207.13482}{{\ttfamily 2207.13482}}.

\bibitem{He:2023qld}
S.~He and Y.~Tang, \emph{{Jumpstarting (elliptic) symbol integrations for loop
  integrals}},  \href{https://arxiv.org/abs/2304.01776}{{\ttfamily
  2304.01776}}.

\bibitem{Arkani-Hamed:2017ahv}
N.~Arkani-Hamed and E.Y.~Yuan, \emph{{One-Loop Integrals from Spherical
  Projections of Planes and Quadrics}},
  \href{https://arxiv.org/abs/1712.09991}{{\ttfamily 1712.09991}}.

\bibitem{Abreu:2017enx}
S.~Abreu, R.~Britto, C.~Duhr and E.~Gardi, \emph{{Algebraic Structure of Cut
  Feynman Integrals and the Diagrammatic Coaction}},
  \href{https://doi.org/10.1103/PhysRevLett.119.051601}{\emph{Phys. Rev. Lett.}
  {\bfseries 119} (2017) 051601}
  [\href{https://arxiv.org/abs/1703.05064}{{\ttfamily 1703.05064}}].

\bibitem{Abreu:2017mtm}
S.~Abreu, R.~Britto, C.~Duhr and E.~Gardi, \emph{{Diagrammatic Hopf algebra of
  cut Feynman integrals: the one-loop case}},
  \href{https://doi.org/10.1007/JHEP12(2017)090}{\emph{JHEP} {\bfseries 12}
  (2017) 090} [\href{https://arxiv.org/abs/1704.07931}{{\ttfamily
  1704.07931}}].

\bibitem{Chen:2022fyw}
J.~Chen, C.~Ma and L.L.~Yang, \emph{{Alphabet of one-loop Feynman integrals
  *}}, \href{https://doi.org/10.1088/1674-1137/ac6e37}{\emph{Chin. Phys. C}
  {\bfseries 46} (2022) 093104}
  [\href{https://arxiv.org/abs/2201.12998}{{\ttfamily 2201.12998}}].

\bibitem{Dlapa:2023cvx}
C.~Dlapa, M.~Helmer, G.~Papathanasiou and F.~Tellander, \emph{{Symbol Alphabets
  from the Landau Singular Locus}},
  \href{https://arxiv.org/abs/2304.02629}{{\ttfamily 2304.02629}}.

\bibitem{Jiang:2024eaj}
X.~Jiang, J.~Liu, X.~Xu and L.L.~Yang, \emph{{Symbol letters of Feynman
  integrals from Gram determinants}},
  \href{https://arxiv.org/abs/2401.07632}{{\ttfamily 2401.07632}}.

\bibitem{Gaiotto:2011dt}
D.~Gaiotto, J.~Maldacena, A.~Sever and P.~Vieira, \emph{{Pulling the straps of
  polygons}}, \href{https://doi.org/10.1007/JHEP12(2011)011}{\emph{JHEP}
  {\bfseries 12} (2011) 011} [\href{https://arxiv.org/abs/1102.0062}{{\ttfamily
  1102.0062}}].

\bibitem{Dixon:2011pw}
L.J.~Dixon, J.M.~Drummond and J.M.~Henn, \emph{{Bootstrapping the three-loop
  hexagon}}, \href{https://doi.org/10.1007/JHEP11(2011)023}{\emph{JHEP}
  {\bfseries 11} (2011) 023} [\href{https://arxiv.org/abs/1108.4461}{{\ttfamily
  1108.4461}}].

\bibitem{Dixon:2011nj}
L.J.~Dixon, J.M.~Drummond and J.M.~Henn, \emph{{Analytic result for the
  two-loop six-point NMHV amplitude in N=4 super Yang-Mills theory}},
  \href{https://doi.org/10.1007/JHEP01(2012)024}{\emph{JHEP} {\bfseries 01}
  (2012) 024} [\href{https://arxiv.org/abs/1111.1704}{{\ttfamily 1111.1704}}].

\bibitem{Brandhuber:2012vm}
A.~Brandhuber, G.~Travaglini and G.~Yang, \emph{{Analytic two-loop form factors
  in N=4 SYM}}, \href{https://doi.org/10.1007/JHEP05(2012)082}{\emph{JHEP}
  {\bfseries 05} (2012) 082} [\href{https://arxiv.org/abs/1201.4170}{{\ttfamily
  1201.4170}}].

\bibitem{Dixon:2013eka}
L.J.~Dixon, J.M.~Drummond, M.~von Hippel and J.~Pennington, \emph{{Hexagon
  functions and the three-loop remainder function}},
  \href{https://doi.org/10.1007/JHEP12(2013)049}{\emph{JHEP} {\bfseries 12}
  (2013) 049} [\href{https://arxiv.org/abs/1308.2276}{{\ttfamily 1308.2276}}].

\bibitem{Dixon:2014voa}
L.J.~Dixon, J.M.~Drummond, C.~Duhr and J.~Pennington, \emph{{The four-loop
  remainder function and multi-Regge behavior at NNLLA in planar N = 4
  super-Yang-Mills theory}},
  \href{https://doi.org/10.1007/JHEP06(2014)116}{\emph{JHEP} {\bfseries 06}
  (2014) 116} [\href{https://arxiv.org/abs/1402.3300}{{\ttfamily 1402.3300}}].

\bibitem{Dixon:2014iba}
L.J.~Dixon and M.~von Hippel, \emph{{Bootstrapping an NMHV amplitude through
  three loops}}, \href{https://doi.org/10.1007/JHEP10(2014)065}{\emph{JHEP}
  {\bfseries 10} (2014) 065} [\href{https://arxiv.org/abs/1408.1505}{{\ttfamily
  1408.1505}}].

\bibitem{Drummond:2014ffa}
J.M.~Drummond, G.~Papathanasiou and M.~Spradlin, \emph{{A Symbol of Uniqueness:
  The Cluster Bootstrap for the 3-Loop MHV Heptagon}},
  \href{https://doi.org/10.1007/JHEP03(2015)072}{\emph{JHEP} {\bfseries 03}
  (2015) 072} [\href{https://arxiv.org/abs/1412.3763}{{\ttfamily 1412.3763}}].

\bibitem{Dixon:2015iva}
L.J.~Dixon, M.~von Hippel and A.J.~McLeod, \emph{{The four-loop six-gluon NMHV
  ratio function}}, \href{https://doi.org/10.1007/JHEP01(2016)053}{\emph{JHEP}
  {\bfseries 01} (2016) 053}
  [\href{https://arxiv.org/abs/1509.08127}{{\ttfamily 1509.08127}}].

\bibitem{Dixon:2016apl}
L.J.~Dixon, M.~von Hippel, A.J.~McLeod and J.~Trnka, \emph{{Multi-loop
  positivity of the planar $ \mathcal{N} $ = 4 SYM six-point amplitude}},
  \href{https://doi.org/10.1007/JHEP02(2017)112}{\emph{JHEP} {\bfseries 02}
  (2017) 112} [\href{https://arxiv.org/abs/1611.08325}{{\ttfamily
  1611.08325}}].

\bibitem{Dixon:2016nkn}
L.J.~Dixon, J.~Drummond, T.~Harrington, A.J.~McLeod, G.~Papathanasiou and
  M.~Spradlin, \emph{{Heptagons from the Steinmann Cluster Bootstrap}},
  \href{https://doi.org/10.1007/JHEP02(2017)137}{\emph{JHEP} {\bfseries 02}
  (2017) 137} [\href{https://arxiv.org/abs/1612.08976}{{\ttfamily
  1612.08976}}].

\bibitem{Li:2016ctv}
Y.~Li and H.X.~Zhu, \emph{{Bootstrapping Rapidity Anomalous Dimensions for
  Transverse-Momentum Resummation}},
  \href{https://doi.org/10.1103/PhysRevLett.118.022004}{\emph{Phys. Rev. Lett.}
  {\bfseries 118} (2017) 022004}
  [\href{https://arxiv.org/abs/1604.01404}{{\ttfamily 1604.01404}}].

\bibitem{Almelid:2017qju}
O.~Almelid, C.~Duhr, E.~Gardi, A.~McLeod and C.D.~White, \emph{{Bootstrapping
  the QCD soft anomalous dimension}},
  \href{https://doi.org/10.1007/JHEP09(2017)073}{\emph{JHEP} {\bfseries 09}
  (2017) 073} [\href{https://arxiv.org/abs/1706.10162}{{\ttfamily
  1706.10162}}].

\bibitem{Chicherin:2017dob}
D.~Chicherin, J.~Henn and V.~Mitev, \emph{{Bootstrapping pentagon functions}},
  \href{https://doi.org/10.1007/JHEP05(2018)164}{\emph{JHEP} {\bfseries 05}
  (2018) 164} [\href{https://arxiv.org/abs/1712.09610}{{\ttfamily
  1712.09610}}].

\bibitem{Henn:2018cdp}
J.~Henn, E.~Herrmann and J.~Parra-Martinez, \emph{{Bootstrapping two-loop
  Feynman integrals for planar $ \mathcal{N}=4 $ sYM}},
  \href{https://doi.org/10.1007/JHEP10(2018)059}{\emph{JHEP} {\bfseries 10}
  (2018) 059} [\href{https://arxiv.org/abs/1806.06072}{{\ttfamily
  1806.06072}}].

\bibitem{Drummond:2018caf}
J.~Drummond, J.~Foster, O.~G\"urdo\u{g}an and G.~Papathanasiou, \emph{{Cluster
  adjacency and the four-loop NMHV heptagon}},
  \href{https://doi.org/10.1007/JHEP03(2019)087}{\emph{JHEP} {\bfseries 03}
  (2019) 087} [\href{https://arxiv.org/abs/1812.04640}{{\ttfamily
  1812.04640}}].

\bibitem{Caron-Huot:2019vjl}
S.~Caron-Huot, L.J.~Dixon, F.~Dulat, M.~von Hippel, A.J.~McLeod and
  G.~Papathanasiou, \emph{{Six-Gluon amplitudes in planar $ \mathcal{N} $ = 4
  super-Yang-Mills theory at six and seven loops}},
  \href{https://doi.org/10.1007/JHEP08(2019)016}{\emph{JHEP} {\bfseries 08}
  (2019) 016} [\href{https://arxiv.org/abs/1903.10890}{{\ttfamily
  1903.10890}}].

\bibitem{Caron-Huot:2020bkp}
S.~Caron-Huot, L.J.~Dixon, J.M.~Drummond, F.~Dulat, J.~Foster,
  O.~G\"urdo\u{g}an et~al., \emph{{The Steinmann Cluster Bootstrap for $N$ = 4
  Super Yang-Mills Amplitudes}},
  \href{https://doi.org/10.22323/1.376.0003}{\emph{PoS} {\bfseries CORFU2019}
  (2020) 003} [\href{https://arxiv.org/abs/2005.06735}{{\ttfamily
  2005.06735}}].

\bibitem{Dixon:2020cnr}
L.J.~Dixon and Y.-T.~Liu, \emph{{Lifting Heptagon Symbols to Functions}},
  \href{https://doi.org/10.1007/JHEP10(2020)031}{\emph{JHEP} {\bfseries 10}
  (2020) 031} [\href{https://arxiv.org/abs/2007.12966}{{\ttfamily
  2007.12966}}].

\bibitem{Dixon:2020bbt}
L.J.~Dixon, A.J.~McLeod and M.~Wilhelm, \emph{{A Three-Point Form Factor
  Through Five Loops}},
  \href{https://doi.org/10.1007/JHEP04(2021)147}{\emph{JHEP} {\bfseries 04}
  (2021) 147} [\href{https://arxiv.org/abs/2012.12286}{{\ttfamily
  2012.12286}}].

\bibitem{Guo:2021bym}
Y.~Guo, L.~Wang and G.~Yang, \emph{{Bootstrapping a Two-Loop Four-Point Form
  Factor}}, \href{https://doi.org/10.1103/PhysRevLett.127.151602}{\emph{Phys.
  Rev. Lett.} {\bfseries 127} (2021) 151602}
  [\href{https://arxiv.org/abs/2106.01374}{{\ttfamily 2106.01374}}].

\bibitem{Dixon:2022rse}
L.J.~Dixon, O.~Gurdogan, A.J.~McLeod and M.~Wilhelm, \emph{{Bootstrapping a
  stress-tensor form factor through eight loops}},
  \href{https://doi.org/10.1007/JHEP07(2022)153}{\emph{JHEP} {\bfseries 07}
  (2022) 153} [\href{https://arxiv.org/abs/2204.11901}{{\ttfamily
  2204.11901}}].

\bibitem{Dixon:2022xqh}
L.J.~Dixon, O.~G\"urdo\u{g}an, Y.-T.~Liu, A.J.~McLeod and M.~Wilhelm,
  \emph{{Antipodal Self-Duality for a Four-Particle Form Factor}},
  \href{https://doi.org/10.1103/PhysRevLett.130.111601}{\emph{Phys. Rev. Lett.}
  {\bfseries 130} (2023) 111601}
  [\href{https://arxiv.org/abs/2212.02410}{{\ttfamily 2212.02410}}].

\bibitem{Chen:2020uyk}
J.~Chen, X.~Jiang, X.~Xu and L.L.~Yang, \emph{{Constructing canonical Feynman
  integrals with intersection theory}},
  \href{https://doi.org/10.1016/j.physletb.2021.136085}{\emph{Phys. Lett. B}
  {\bfseries 814} (2021) 136085}
  [\href{https://arxiv.org/abs/2008.03045}{{\ttfamily 2008.03045}}].

\bibitem{Chen:2022lzr}
J.~Chen, X.~Jiang, C.~Ma, X.~Xu and L.L.~Yang, \emph{{Baikov representations,
  intersection theory, and canonical Feynman integrals}},
  \href{https://doi.org/10.1007/JHEP07(2022)066}{\emph{JHEP} {\bfseries 07}
  (2022) 066} [\href{https://arxiv.org/abs/2202.08127}{{\ttfamily
  2202.08127}}].

\bibitem{Chicherin:2018old}
D.~Chicherin, T.~Gehrmann, J.M.~Henn, P.~Wasser, Y.~Zhang and S.~Zoia,
  \emph{{All Master Integrals for Three-Jet Production at
  Next-to-Next-to-Leading Order}},
  \href{https://doi.org/10.1103/PhysRevLett.123.041603}{\emph{Phys. Rev. Lett.}
  {\bfseries 123} (2019) 041603}
  [\href{https://arxiv.org/abs/1812.11160}{{\ttfamily 1812.11160}}].

\bibitem{Bern:2014kca}
Z.~Bern, E.~Herrmann, S.~Litsey, J.~Stankowicz and J.~Trnka, \emph{{Logarithmic
  Singularities and Maximally Supersymmetric Amplitudes}},
  \href{https://doi.org/10.1007/JHEP06(2015)202}{\emph{JHEP} {\bfseries 06}
  (2015) 202} [\href{https://arxiv.org/abs/1412.8584}{{\ttfamily 1412.8584}}].

\bibitem{Henn:2020lye}
J.~Henn, B.~Mistlberger, V.A.~Smirnov and P.~Wasser, \emph{{Constructing d-log
  integrands and computing master integrals for three-loop four-particle
  scattering}}, \href{https://doi.org/10.1007/JHEP04(2020)167}{\emph{JHEP}
  {\bfseries 04} (2020) 167}
  [\href{https://arxiv.org/abs/2002.09492}{{\ttfamily 2002.09492}}].

\bibitem{Dlapa:2021qsl}
C.~Dlapa, X.~Li and Y.~Zhang, \emph{{Leading singularities in Baikov
  representation and Feynman integrals with uniform transcendental weight}},
  \href{https://doi.org/10.1007/JHEP07(2021)227}{\emph{JHEP} {\bfseries 07}
  (2021) 227} [\href{https://arxiv.org/abs/2103.04638}{{\ttfamily
  2103.04638}}].

\bibitem{Arkani-Hamed:2010pyv}
N.~Arkani-Hamed, J.L.~Bourjaily, F.~Cachazo and J.~Trnka, \emph{{Local
  Integrals for Planar Scattering Amplitudes}},
  \href{https://doi.org/10.1007/JHEP06(2012)125}{\emph{JHEP} {\bfseries 06}
  (2012) 125} [\href{https://arxiv.org/abs/1012.6032}{{\ttfamily 1012.6032}}].

\bibitem{Arkani-Hamed:2012zlh}
N.~Arkani-Hamed, J.L.~Bourjaily, F.~Cachazo, A.B.~Goncharov, A.~Postnikov and
  J.~Trnka, \emph{{Grassmannian Geometry of Scattering Amplitudes}}, Cambridge
  University Press (4, 2016),
  \href{https://doi.org/10.1017/CBO9781316091548}{10.1017/CBO9781316091548},
  [\href{https://arxiv.org/abs/1212.5605}{{\ttfamily 1212.5605}}].

\bibitem{Dlapa:2020cwj}
C.~Dlapa, J.~Henn and K.~Yan, \emph{{Deriving canonical differential equations
  for Feynman integrals from a single uniform weight integral}},
  \href{https://doi.org/10.1007/JHEP05(2020)025}{\emph{JHEP} {\bfseries 05}
  (2020) 025} [\href{https://arxiv.org/abs/2002.02340}{{\ttfamily
  2002.02340}}].

\bibitem{Lee:2020zfb}
R.N.~Lee, \emph{{Libra: A package for transformation of differential systems
  for multiloop integrals}},
  \href{https://doi.org/10.1016/j.cpc.2021.108058}{\emph{Comput. Phys. Commun.}
  {\bfseries 267} (2021) 108058}
  [\href{https://arxiv.org/abs/2012.00279}{{\ttfamily 2012.00279}}].

\bibitem{Prausa:2017ltv}
M.~Prausa, \emph{{epsilon: A tool to find a canonical basis of master
  integrals}}, \href{https://doi.org/10.1016/j.cpc.2017.05.026}{\emph{Comput.
  Phys. Commun.} {\bfseries 219} (2017) 361}
  [\href{https://arxiv.org/abs/1701.00725}{{\ttfamily 1701.00725}}].

\bibitem{Gituliar:2017vzm}
O.~Gituliar and V.~Magerya, \emph{{Fuchsia: a tool for reducing differential
  equations for Feynman master integrals to epsilon form}},
  \href{https://doi.org/10.1016/j.cpc.2017.05.004}{\emph{Comput. Phys. Commun.}
  {\bfseries 219} (2017) 329}
  [\href{https://arxiv.org/abs/1701.04269}{{\ttfamily 1701.04269}}].

\bibitem{Meyer:2017joq}
C.~Meyer, \emph{{Algorithmic transformation of multi-loop master integrals to a
  canonical basis with CANONICA}},
  \href{https://doi.org/10.1016/j.cpc.2017.09.014}{\emph{Comput. Phys. Commun.}
  {\bfseries 222} (2018) 295}
  [\href{https://arxiv.org/abs/1705.06252}{{\ttfamily 1705.06252}}].

\bibitem{Meyer:2016slj}
C.~Meyer, \emph{{Transforming differential equations of multi-loop Feynman
  integrals into canonical form}},
  \href{https://doi.org/10.1007/JHEP04(2017)006}{\emph{JHEP} {\bfseries 04}
  (2017) 006} [\href{https://arxiv.org/abs/1611.01087}{{\ttfamily
  1611.01087}}].

\bibitem{Baikov:1996iu}
P.A.~Baikov, \emph{{Explicit solutions of the multiloop integral recurrence
  relations and its application}},
  \href{https://doi.org/10.1016/S0168-9002(97)00126-5}{\emph{Nucl. Instrum.
  Meth. A} {\bfseries 389} (1997) 347}
  [\href{https://arxiv.org/abs/hep-ph/9611449}{{\ttfamily hep-ph/9611449}}].

\bibitem{Mastrolia:2018uzb}
P.~Mastrolia and S.~Mizera, \emph{{Feynman Integrals and Intersection Theory}},
  \href{https://doi.org/10.1007/JHEP02(2019)139}{\emph{JHEP} {\bfseries 02}
  (2019) 139} [\href{https://arxiv.org/abs/1810.03818}{{\ttfamily
  1810.03818}}].

\bibitem{Frellesvig:2019uqt}
H.~Frellesvig, F.~Gasparotto, M.K.~Mandal, P.~Mastrolia, L.~Mattiazzi and
  S.~Mizera, \emph{{Vector Space of Feynman Integrals and Multivariate
  Intersection Numbers}},
  \href{https://doi.org/10.1103/PhysRevLett.123.201602}{\emph{Phys. Rev. Lett.}
  {\bfseries 123} (2019) 201602}
  [\href{https://arxiv.org/abs/1907.02000}{{\ttfamily 1907.02000}}].

\bibitem{Mizera:2019ose}
S.~Mizera, \emph{{Status of Intersection Theory and Feynman Integrals}},
  \href{https://doi.org/10.22323/1.383.0016}{\emph{PoS} {\bfseries MA2019}
  (2019) 016} [\href{https://arxiv.org/abs/2002.10476}{{\ttfamily
  2002.10476}}].

\bibitem{Matsubara-Heo:2020lzo}
S.-J.~Matsubara-Heo, \emph{{Computing cohomology intersection numbers of GKZ
  hypergeometric systems}},
  \href{https://doi.org/10.22323/1.383.0013}{\emph{PoS} {\bfseries MA2019}
  (2022) 013} [\href{https://arxiv.org/abs/2008.03176}{{\ttfamily
  2008.03176}}].

\bibitem{Chestnov:2022xsy}
V.~Chestnov, H.~Frellesvig, F.~Gasparotto, M.K.~Mandal and P.~Mastrolia,
  \emph{{Intersection Numbers from Higher-order Partial Differential
  Equations}},  \href{https://arxiv.org/abs/2209.01997}{{\ttfamily
  2209.01997}}.

\bibitem{Weinzierl:2020xyy}
S.~Weinzierl, \emph{{On the computation of intersection numbers for twisted
  cocycles}}, \href{https://doi.org/10.1063/5.0054292}{\emph{J. Math. Phys.}
  {\bfseries 62} (2021) 072301}
  [\href{https://arxiv.org/abs/2002.01930}{{\ttfamily 2002.01930}}].

\bibitem{Jiang:2023oyq}
X.~Jiang, M.~Lian and L.L.~Yang, \emph{{Recursive structure of Baikov
  representations: The top-down reduction with intersection theory}},
  \href{https://doi.org/10.1103/PhysRevD.109.076020}{\emph{Phys. Rev. D}
  {\bfseries 109} (2024) 076020}
  [\href{https://arxiv.org/abs/2312.03453}{{\ttfamily 2312.03453}}].

\bibitem{Jiang:2023qnl}
X.~Jiang and L.L.~Yang, \emph{{Recursive structure of Baikov representations:
  Generics and application to symbology}},
  \href{https://doi.org/10.1103/PhysRevD.108.076004}{\emph{Phys. Rev. D}
  {\bfseries 108} (2023) 076004}
  [\href{https://arxiv.org/abs/2303.11657}{{\ttfamily 2303.11657}}].

\bibitem{Crisanti:2024onv}
G.~Crisanti and S.~Smith, \emph{{Feynman Integral Reductions by Intersection
  Theory with Orthogonal Bases and Closed Formulae}},
  \href{https://arxiv.org/abs/2405.18178}{{\ttfamily 2405.18178}}.

\bibitem{Frellesvig:2020qot}
H.~Frellesvig, F.~Gasparotto, S.~Laporta, M.K.~Mandal, P.~Mastrolia,
  L.~Mattiazzi et~al., \emph{{Decomposition of Feynman Integrals by
  Multivariate Intersection Numbers}},
  \href{https://doi.org/10.1007/JHEP03(2021)027}{\emph{JHEP} {\bfseries 03}
  (2021) 027} [\href{https://arxiv.org/abs/2008.04823}{{\ttfamily
  2008.04823}}].

\bibitem{Mizera:2019vvs}
S.~Mizera and A.~Pokraka, \emph{{From Infinity to Four Dimensions: Higher
  Residue Pairings and Feynman Integrals}},
  \href{https://doi.org/10.1007/JHEP02(2020)159}{\emph{JHEP} {\bfseries 02}
  (2020) 159} [\href{https://arxiv.org/abs/1910.11852}{{\ttfamily
  1910.11852}}].

\bibitem{Chestnov:2022alh}
V.~Chestnov, F.~Gasparotto, M.K.~Mandal, P.~Mastrolia, S.J.~Matsubara-Heo,
  H.J.~Munch et~al., \emph{{Macaulay matrix for Feynman integrals: linear
  relations and intersection numbers}},
  \href{https://doi.org/10.1007/JHEP09(2022)187}{\emph{JHEP} {\bfseries 09}
  (2022) 187} [\href{https://arxiv.org/abs/2204.12983}{{\ttfamily
  2204.12983}}].

\bibitem{Caron-Huot:2021xqj}
S.~Caron-Huot and A.~Pokraka, \emph{{Duals of Feynman integrals. Part I.
  Differential equations}},
  \href{https://doi.org/10.1007/JHEP12(2021)045}{\emph{JHEP} {\bfseries 12}
  (2021) 045} [\href{https://arxiv.org/abs/2104.06898}{{\ttfamily
  2104.06898}}].

\bibitem{Caron-Huot:2021iev}
S.~Caron-Huot and A.~Pokraka, \emph{{Duals of Feynman Integrals. Part II.
  Generalized unitarity}},
  \href{https://doi.org/10.1007/JHEP04(2022)078}{\emph{JHEP} {\bfseries 04}
  (2022) 078} [\href{https://arxiv.org/abs/2112.00055}{{\ttfamily
  2112.00055}}].

\bibitem{Giroux:2022wav}
M.~Giroux and A.~Pokraka, \emph{{Loop-by-loop differential equations for dual
  (elliptic) Feynman integrals}},
  \href{https://doi.org/10.1007/JHEP03(2023)155}{\emph{JHEP} {\bfseries 03}
  (2023) 155} [\href{https://arxiv.org/abs/2210.09898}{{\ttfamily
  2210.09898}}].

\bibitem{De:2023xue}
S.~De and A.~Pokraka, \emph{{Cosmology meets cohomology}},
  \href{https://doi.org/10.1007/JHEP03(2024)156}{\emph{JHEP} {\bfseries 03}
  (2024) 156} [\href{https://arxiv.org/abs/2308.03753}{{\ttfamily
  2308.03753}}].

\bibitem{Duhr:2024rxe}
C.~Duhr, F.~Porkert, C.~Semper and S.F.~Stawinski, \emph{{Twisted Riemann
  bilinear relations and Feynman integrals}},
  \href{https://arxiv.org/abs/2407.17175}{{\ttfamily 2407.17175}}.

\bibitem{Chen:2023kgw}
J.~Chen, B.~Feng and L.L.~Yang, \emph{{Intersection theory rules symbology}},
  \href{https://doi.org/10.1007/s11433-023-2239-8}{\emph{Sci. China Phys. Mech.
  Astron.} {\bfseries 67} (2024) 221011}
  [\href{https://arxiv.org/abs/2305.01283}{{\ttfamily 2305.01283}}].

\bibitem{Brunello:2024tqf}
G.~Brunello, V.~Chestnov and P.~Mastrolia, \emph{{Intersection Numbers from
  Companion Tensor Algebra}},
  \href{https://arxiv.org/abs/2408.16668}{{\ttfamily 2408.16668}}.

\bibitem{Zhang:2016kfo}
Y.~Zhang, \emph{{Lecture Notes on Multi-loop Integral Reduction and Applied
  Algebraic Geometry}},  12, 2016
  [\href{https://arxiv.org/abs/1612.02249}{{\ttfamily 1612.02249}}].

\bibitem{Larsen:2017aqb}
K.J.~Larsen and R.~Rietkerk, \emph{{MultivariateResidues: a Mathematica package
  for computing multivariate residues}},
  \href{https://doi.org/10.1016/j.cpc.2017.08.025}{\emph{Comput. Phys. Commun.}
  {\bfseries 222} (2018) 250}
  [\href{https://arxiv.org/abs/1701.01040}{{\ttfamily 1701.01040}}].

\bibitem{Binoth:2000ps}
T.~Binoth and G.~Heinrich, \emph{{An automatized algorithm to compute infrared
  divergent multiloop integrals}},
  \href{https://doi.org/10.1016/S0550-3213(00)00429-6}{\emph{Nucl. Phys. B}
  {\bfseries 585} (2000) 741}
  [\href{https://arxiv.org/abs/hep-ph/0004013}{{\ttfamily hep-ph/0004013}}].

\bibitem{Binoth:2003ak}
T.~Binoth and G.~Heinrich, \emph{{Numerical evaluation of multiloop integrals
  by sector decomposition}},
  \href{https://doi.org/10.1016/j.nuclphysb.2003.12.023}{\emph{Nucl. Phys. B}
  {\bfseries 680} (2004) 375}
  [\href{https://arxiv.org/abs/hep-ph/0305234}{{\ttfamily hep-ph/0305234}}].

\bibitem{Binoth:2004jv}
T.~Binoth and G.~Heinrich, \emph{{Numerical evaluation of phase space integrals
  by sector decomposition}},
  \href{https://doi.org/10.1016/j.nuclphysb.2004.06.005}{\emph{Nucl. Phys. B}
  {\bfseries 693} (2004) 134}
  [\href{https://arxiv.org/abs/hep-ph/0402265}{{\ttfamily hep-ph/0402265}}].

\bibitem{Heinrich:2008si}
G.~Heinrich, \emph{{Sector Decomposition}},
  \href{https://doi.org/10.1142/S0217751X08040263}{\emph{Int. J. Mod. Phys. A}
  {\bfseries 23} (2008) 1457}
  [\href{https://arxiv.org/abs/0803.4177}{{\ttfamily 0803.4177}}].

\bibitem{Johansson:2015ava}
H.~Johansson, D.A.~Kosower, K.J.~Larsen and M.~S\o{}gaard, \emph{{Cross-Order
  Integral Relations from Maximal Cuts}},
  \href{https://doi.org/10.1103/PhysRevD.92.025015}{\emph{Phys. Rev. D}
  {\bfseries 92} (2015) 025015}
  [\href{https://arxiv.org/abs/1503.06711}{{\ttfamily 1503.06711}}].

\bibitem{Arkani-Hamed:2009ljj}
N.~Arkani-Hamed, F.~Cachazo, C.~Cheung and J.~Kaplan, \emph{{A Duality For The
  S Matrix}}, \href{https://doi.org/10.1007/JHEP03(2010)020}{\emph{JHEP}
  {\bfseries 03} (2010) 020} [\href{https://arxiv.org/abs/0907.5418}{{\ttfamily
  0907.5418}}].

\bibitem{tsikh1992multidimensional}
A.K.~Tsikh, \emph{Multidimensional residues and their applications}, vol.~103,
  Citeseer (1992).

\bibitem{auizenberg1983integral}
I.A.~Aizenberg and A.P.~Yuzhakov, \emph{Integral representations and residues
  in multidimensional complex analysis}, vol.~58, American Mathematical Soc.
  (1983).

\end{thebibliography}\endgroup








\end{document}